\documentclass{article}
\usepackage{graphicx} 
\usepackage{subfigure} 
\usepackage{algorithm}
\usepackage{algorithmic}
\usepackage{amsmath}
\usepackage{amsthm}
\usepackage{amsfonts}
\usepackage{amssymb}
\usepackage{multirow}

\def\argmin{\mathop{\mathrm{argmin}}}

\usepackage{times}
\usepackage{helvet}
\usepackage{courier}
\usepackage[numbers]{natbib}

\title{The Impact of Visual Appearance on User Response in Online Display Advertising}

\author{     Javad Azimi \\ azimi@eecs.oregonstate.edu\\ Oregon State University \and
  Ruofei Zhang \\ rzhang @yahoo-inc.com\\  Yahoo! Labs, Silicon Valley \and 
 Yang Zhou \\ yangzhou @yahoo-inc.com\\  Yahoo! Labs, Silicon Valley \and 
  Vidhya Navalpakkam \\ nvidhya@yahoo-inc.com \\  Yahoo! Labs, Silicon Valley \and 
  Jianchang Mao \\ jmao@yahoo-inc.com\\  Yahoo! Labs, Silicon Valley \and
 Xiaoli Fern \\ xfern@eecs.oregonstate.edu\\  Oregon State University\\
}

\begin{document} 

\maketitle
\begin{abstract}
Display advertising has been a significant source of revenue for publishers and ad networks in online advertising ecosystem. One of the main goals in display advertising is to maximize user response rate for advertising campaigns, such as click through rates (CTR) or conversion rates. Although in the online advertising industry we believe that the visual appearance of ads (creatives) matters for propensity of user response, there is no published work so far to address this topic via a systematic data-driven approach. In this paper we quantitatively study the relationship between the visual appearance and performance of creatives using large scale data in the world's largest display ads exchange system, RightMedia. We designed a set of $43$ visual features, some of which are novel and some are inspired by related work. We extracted these features from real creatives served on RightMedia. We also designed and conducted a series of experiments to evaluate the effectiveness of visual features for CTR prediction, ranking and performance classification. Based on the evaluation results, we selected a subset of features that have the most important impact on CTR. We believe that the findings presented in this paper will be very useful for the online advertising industry in designing high-performance creatives. It also provides the research community with the first ever data set, initial insights into visual appearance's effect on user response propensity, and evaluation benchmarks for further study.
\end{abstract}

\section{Introduction}
\label{sec:introduction}

The Internet revolution has transformed how people experience information, media and advertising. Web advertising, although nonexisting twenty years ago, has become a vital component of the modern Internet, where advertisements are delivered from advertisers to users through different online channels. Recent trends have shown that an increasingly large share of advertisers' budgets are devoted to the online world, and online advertising spending has greatly outpaced some of the traditional advertising media, such as radio and magazine. Display advertising is one type of online advertising which, together with search advertising, contributes the majority of the revenue for many large Internet companies. In display advertising, display ad instances are shown to the user on webpages in different formats such as image, flash, and video. Each display ad instance is called a creative. By showing the creatives, advertisers aim to either promote brand awareness among users (brand advertising) or receive desirable responses from users (performance advertising), such as the action of purchasing, clicking or signing up for a promotion list from the advertiser's website. In performance advertising, the advertiser strives to optimize their ad's performance metrics such as the effective cost per click (eCPC) or effective cost per action (eCPA), which in turn relates to maximizing the user response rate on the creatives as measured by click through rates (CTR) or conversion rates (CVR). There are several factors that greatly influence the user response rate of display advertising campaigns: 1) the position of the ads on the webpage; 2) the relevancy of the ads to the online users, which is generally captured by the targeting profiles of the advertising campaigns; 3) the relevancy of the ads to the webpage content and 4) the quality and visual appearance of the creatives.


The problem of predicting the user response rate for online ads, especially CTR, has been studied by several researchers in the last few years. One major research focus has been in predicting clicks by studying the relationship between CTR and the aforementioned ad factors (and their combinations). For example in \citep{Chakrabarti08}, the authors considered the ad's relevancy to the content of the webpage in predicting CTR. They show that improving the ad's content relevancy is more efficient than considering the content of ads by themselves \citep{Richardson07}. Although it is generally believed that visually appealing ads can perform better in attracting online users, as a result of which advertisers always care about the creative designs, there is no, to the best of our knowledge, published work so far to quantitatively study the effect of visual appearance of creatives on campaign performance in online display advertising. This motivates us to investigate the correlation between the visual features of the creative and CTR, regardless of other ad factors, and to predict creative performance based on its visual appearance alone.


Our proposed approach consists of two main steps, 1) feature extraction and 2) correlation investigation. We first extract some informative visual features from the creatives. We introduce $43$ visual features classified into three categories, 1) \emph{global features} which characterize the overall properties of a given creative, 2) \emph{local features} representing the properties of specific parts within a given creative and 3) \emph{advanced features} which are a group of features developed based on more complicated algorithms such as the number of faces and number of characters in a creative. We then develop three regression approaches to predict the CTR based on these features. The study is conducted using real creatives and their performance data from the world's largest display ads exchange system, RightMedia. Based on the weights of developed features, we further select a subset of features that have high impact on the creative's CTR. The benefit of this work is three-fold. First, our findings on the visual features and their relationship to CTR can provide useful recommendations to designers on what features to consider while designing creatives, and/or can help in automated creative generation. Second, the visual features and the regression methods developed here can be used in addition to the traditionally investigated ad factors (such as ad relevancy, position etc.) for improving CTR prediction in online ads selection. Third, it provides the research community with the first ever data set, initial insights into the effect of visual appearance on user response propensity, and evaluation benchmarks for further study.

The paper is organized as follows. Section \ref{sec:related-works} introduces the related work. We introduce the visual features in Section \ref{sec:feature-extraction}. The regression and feature selection results for CTR prediction are presented in Section \ref{sec:experimental-results}, followed by our conclusion in Section \ref{sec:conclusion}.

\section{Background and Related Work}
\label{sec:related-works}
%
The relationship between various print ad characteristics and measures of advertising effectiveness has been studied by advertising researchers for almost a century. A wide variety of  characteristics have been investigated. These characteristics are roughly in two categories: mechanical and content-based. The mechanical characteristics include ad size, number of colors, proportional of illustrations to copy, the absence of borders, and type size. The content factors include message appeal like status, quality, fear and fantasy, attention-getting techniques like free offers, presence of women, and psycholinguistic variables like product or personal reference in headline, interrogative or imperative headline, visual rhetorics, among others. See~\citep{mcquarrie1999visual} for summaries.

Even though the online advertising has taken a large market share of the advertising industry, and the whole industry is steadily and continuously shifting to the online domain, study on the effectiveness of the counterpart of print ads online, generally called display ads, is limited. We list the studies of several factors below.

Some existing studies try to investigate the effect of several different factors on the performance of display advertising campaigns. These factors include targeting and obtrusiveness~\citep{goldfarb2010online}, advertisement size (large vs. small) and ad exposure format (intrusive vs. voluntary)~\citep{chatterjee2008unclicked}, cognitive impact from ad size and animation~\citep{li1999cognitive}, emotional appeal and incentive offering in the ads~\citep{donthu2004emotional}, repetition of varied execution vs. single execution~\citep{yaveroglu2008advertising}.

To the best of our knowledge, we are not aware of any study on the relationship between the visual appearance and the performance of creatives in online display ads. We try to tackle this problem by first defining a set of visual features and then evaluating their effects on ad performance, specifically CTR in our experiments, from the actively served ad campaigns on the world's largest ad exchange system, RightMedia. Below, we present some previous computational studies on image properties which provide us inspiration in designing our visual features.

 There are several studies that try to investigate a specific property of images (photos or paintings) using computational approaches. Such properties include quality and aesthetic in photos~\citep{Luo08, Ke06,Tong04,Datta06} or in paintings~\citep{Li09}, saliency~\citep{Itti98}, composition~\citep{Gooch01,Renjie10}, color harmony~\citep{Cohen06} and memorability~\citep{Isola2011}.

Initial work on image quality evaluation concentrated on evaluating and reconstructing low graded, compressed or degraded images by simple noise model~\citep{Damera00,Bovik05}. However, in most of the beauty evaluation work, including this paper, we assume that high quality images are available and we are interested in evaluating the visual aesthetic of images based on visual features.

Recently some researchers tried to evaluate the beauty of an image based on its visual features. 
In~\citep{Ke06} the authors aim to classify the pictures into professional and snapshot photos using some basic features including spatial distribution of edges, color distribution and hue count, etc. In~\citep{Datta06} the authors introduced a regression based approach for rating photos based on their beauty, using features such as average pixel density, colorfulness, saturation hue, and the rule of thirds. In addition to these studies, in \citep{Luo08} the author proposed an approach to classifying images into high and low quality. The main idea comes from the fact that a professional photographer makes the background blurry and the subject distinguishable in the image. By separating the blurry part of the image from the subject, they design a set of well-motivated features from both the subject and the whole image such as the clarity contrast of the subject, lighting, simplicity, color harmony and composition geometry. They show that the combination of these features can provide a promising performance. All of the above work tries to extract visual features from photos. Recently Li et al.~\citep{Li09} tried to extract some features from paintings to evaluate their beauty and classify them into high and low quality. They introduced a set of global and local features, $40$ in total, to capture the painting properties such as the brightness contrast between segments, the brightness contrast across the whole image and the average saturation for the largest segment of the image.

Computational approaches have also been used to investigate other visual properties of an image. In \citep{Isola2011} the authors studied what properties of images make them more memorable. They found that statistical properties of an image such as mean hue, mean saturation, intensity mean, intensity variance, intensity skewness and number of objects do not have any non-trivial correlation with memorability in their generated data set. However, they found that if they label the objects and scenes in the images, they can find a non-trivial and interesting correlation between images and their memorability. For example, their results show that the attendance of human being, close up objects and human scale objects in an image improve its memorability more than natural scene. This result is not possible to be applied to our work since it requires large amounts of supervision to tag different parts of the images. However, we evaluate the impact of the number of human faces in an image in our work.


Color harmonization is another approach for making an image more appealing. In \citep{Cohen06} the authors proposed to harmonize the colors in a given image using harmonization templates from \citep{Mastuda95, Tokumaru02}, which include $8$ different harmonized color templates. We also used color harmony models to evaluate the hue distribution of an image in our experiments.

In summary,  existing work in related areas has focused primarily on properties of an image, photo or painting. In contrast, we examine creatives in online display ads, which contain both graphical features and text. In addition, some of the existing approaches require significant amount of supervision in their feature extraction step, which is not possible in large scale applications where we need to learn from large data sets with minimum amount of supervision. Finally, we would also like to extract a set of features that are visually understandable and can be practically controlled to guide the human designers or automatic creative generators (like in smart ads) to produce high-performance creatives. These objectives make our problem novel and interesting for the online advertising industry.

\section{Feature Extraction}
\label{sec:feature-extraction}
In this section we introduce a set of $43$ different visual features. We categorize the developed features into three different sets, 1) global features, 2) local features and 3) advanced features. A complete list of the features can be found in Table~\ref{table:fs-results}. Below we describe the detailed definition of the proposed features in each category.

In the following sections we use $I$ to indicate an image and use $|I|$ to indicate the size of the image measured by the number of pixels. We use variable $x$ to denote an arbitrary pixel when we do not care about its location in the image. Otherwise we use $(i,j)$ to denote the pixel in the $i$-th row and $j$-th column in the image.

\subsection{Global Features}
\label{ss:gf}
Global features are a set of features which represent the overall properties of the whole image. We describe the details of $19$ different global features in this section.
\vspace{-0.05in}
\subsubsection { Gray Level Features}
\label{ss:gl-contrast}

We describe 3 features extracted from the gray level histogram of the image, namely the gray level contrast $f_1$, number of dominant gray level bins $f_2$, and the standard deviation of the gray level values among all pixels $f_3$.

The gray level contrast is the width of the middle $95\%$ mass in the gray level histogram~\cite{Ke06}. From the original gray level histogram, we prune the extreme $2.5\%$ from the 0 side and $2.5\%$ from the 255 side. Gray level contrast feature $f_1$ is calculated as the width of the remaining histogram.

We count the number of dominant bins in the gray level histogram as our second feature. Suppose the set $G=\{g_0, g_1, \cdots, g_{255}\}$ indicates the set of $256$ bins in the gray level histogram such that $g_i$ is the number of pixel in $i$-th bins. We define the number of dominant gray level bins as
$f_2 = \sum_{k=0}^{255} \textbf{1}(g_k \geq c_1 \max_i g_i)$, where $\textbf{1}(\cdot)$ is the indicator function and $c_1$ is a threshold value which is set to be $0.01$ in this paper. \footnote{This parameter, and similar ones in the rest of the paper, is set inspired by related works such as \cite{Luo08}.}


The last gray level feature, $f_3$, is defined as the standard deviation of gray level values of all pixels in the image. It is used to capture the variance of the gray level distribution.

\subsubsection{Color Distribution}
\label{ss:color-dist}

To avoid distraction from objects in the background, professional photographers tend to keep the background simple. In \cite{Luo08}, the authors use the color distribution of the background to measure this simplicity. We use a similar approach to measure the simplicity of color distribution in the image. For a given image, we quantize each RGB channel into $8$ values, creating a histogram $H_{rgb}=\{h_0,h_1,\cdots,h_{511}\}$ of 512 bins, where $h_i$ indicates the number of pixels in $i$-th bin. We define feature $f_4$ to indicate the number of dominant colors as $f_4 = \sum_{k=0}^{512} \textbf{1} (h_k \geq c_2 \max_i h_i)$ where $c_2=0.01$ is the threshold parameter. 
We also calculate the size of the dominant bin relative to the image size as $
f_5 = \frac{\max_i h_i}{|I|} $. This feature indicates the extent to which one of $512$ colors is dominant in the image.

By replacing the RGB color map with HSV (Hue, Saturation, Value) color map and using the above methods in calculating features $f_4$ and $f_5$, we obtain two other features $f_6$ and $f_7$.

\subsubsection{Model-Based Color Harmony}
\label{ss-model-based-color-harmony}

\begin{figure}[h!]
  \centering
    \includegraphics[width=2.9in, height=1.5in] {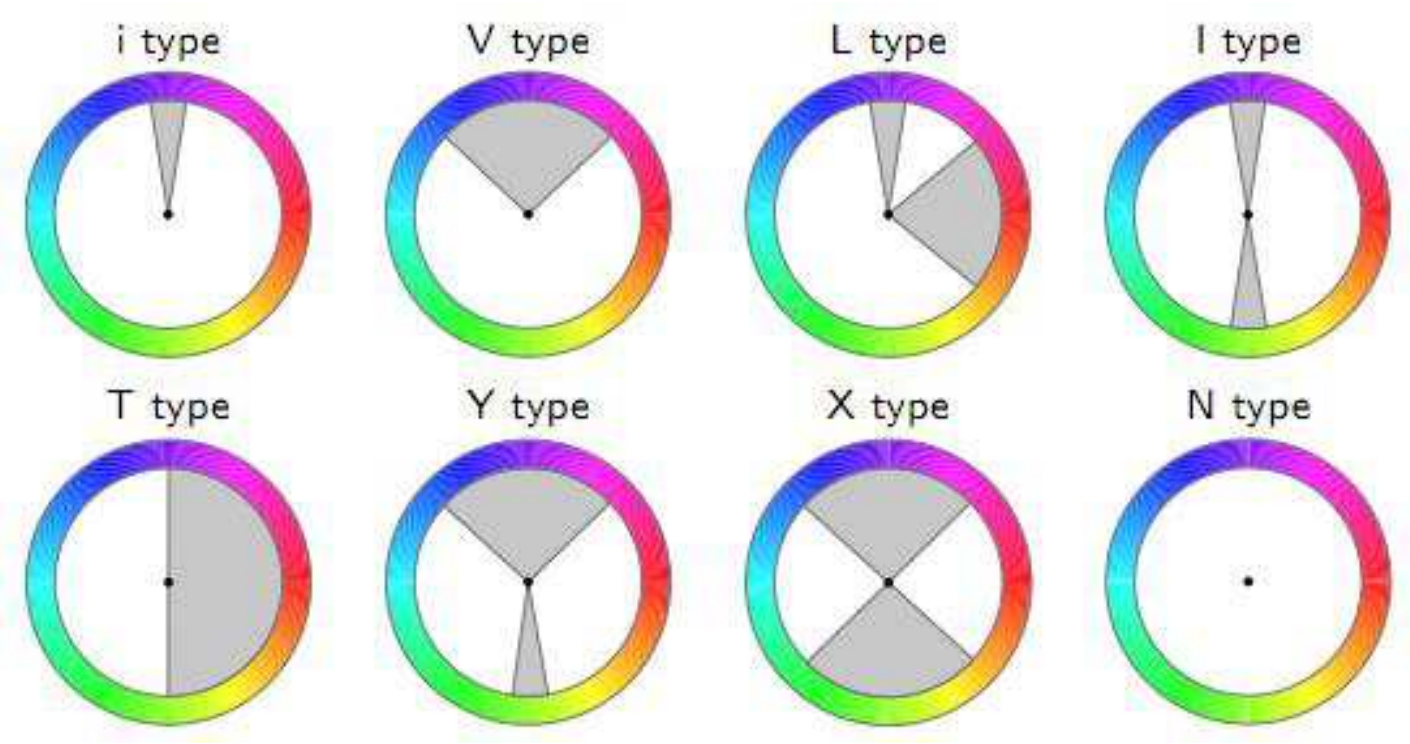}
  \caption{Color harmony models}
\label{fiq:hormony-models}
\end{figure}

The concept of color harmony in this paper is based on $8$ different harmonic color distributions (illustrated in Figure \ref{fiq:hormony-models}) that are based on the hue of the HSV color wheel \cite{Mastuda95}. These  distributions are called \emph{i, V, L, I, T, Y, X, N}. Note that each distribution can be rotated by $0 \leq \alpha \leq 360$ degrees. The specific size of color harmony distributions are set as follows: the large sectors of types $V, Y$ and $X$ are $26\%$ of the disk ($93.6^\circ$); the small sectors of types $i,L,I$ and $Y$ are $5\%$ of the disk ($18^\circ$); the largest sector of type $L$ is $22\%$ of the disk ($79.2\%$); the sector of type $T$ is $50\%$ of the disk ($180^\circ$). The angle between the centers of the two sectors is $180^\circ$ for $I$, $X$, $Y$,  and $90^\circ$ for $L$.

Let us define the set of $8$ distributions as  $\mathcal{D}=\{ d^1,d^2,\cdot\cdot\cdot, d^8 \}$. We say  $\phi (d^i_\alpha,x)$ indicates the hue of the closest point in the $i$-th distribution to $x$ after $\alpha$ degree rotation, where $x$ is any arbitrary pixel in the image. We compute the distance between the hue distribution of our image $I$ and the distribution $d^i \in\mathcal{D}$ as:
\begin{equation}
\gamma(I,d^i)= \argmin_\alpha \frac{1}{|I|}\sum_{x \in I} \parallel hue(x)-\phi(d^i_\alpha,x) \parallel \cdot sat(x),
\end{equation}
where $hue(x)$ and $sat(x)$ indicate the hue and saturation at pixel $x$, and $\|\cdot\|$ denotes the arc-length distance. We are interested in the best fitting model $d^*$ which has the least $\gamma(\cdot)$ value, $ d^*=\argmin_{d^i} \gamma(I,d^i)$. We define feature $f_8=\gamma(I,d^*)$. Intuitively, it tells us how different is the hue distribution of image $I$ from the best fitting model of color harmony.



Some models are superset of other models in Figure \ref{fiq:hormony-models} concluding that the $\gamma(\cdot)$ value of some smaller models are higher than some larger models given any image $I$, e.g. $\gamma(I,d_i)\geq \gamma(I, d_V) \geq \gamma (I,d_T)$. Therefore, if an image hue distribution fits into some small models, type $i,V,L,I$, it fits into larger models as well. This can emphasize the color harmony property of the images which can fit into a few models rather than just one model. We consider this property as one potential positive property of the image. To quantify this property, we introduce a new feature, $f_9$, which indicates the average color harmony deviation from the best two fitted models given an image $I$. In general, in addition to the deviation from the best fitted model illustrated by feature $f_8$, we consider the deviation from the second best fitted model as well, and the average of these two deviations is returned as $f_9$. Clearly, for the images fitting into small color harmony models, we will have $f_8$ and $f_9$ very close to each other. However, for the images which fit into the largest model, we will have $f_9$ considerably larger than $f_8$. We believe these two numerical features can represent the color harmony property of an image appropriately.


\vspace{-0.1in}
\subsubsection{ Color Coherence}
\label{ss:color-coherence}

We extract a set of features based on the color coherence of pixels resulting in connected coherent components~\cite{Pass97}. A connected coherent component in an image is defined as:

\begin{itemize}
	\item A set of pixels that fall into the same bin in the histogram.
	\item For any two pixels $p_i$ and $p_j$ in a connected coherent component $P=\{p_1,p_2,\cdots,p_m\}$ of $m$ pixels, there is a path of sequential  pixels, $p_i, p_{i+1}, \cdot\cdot\cdot, p_j$. Two sequential pixels in a path must be one of the $8$ neighborhoods of each other.
	\item The size of the connected coherent component is larger than a predefined threshold $c_4$. In our experiment we set $c_4=0.01|I|$.
\end{itemize}
\vspace{-0.15in}

We denote the set of connected coherent components and their color index as $\mathcal{P}=\{(P_1,h_1), (P_2,h_2), \cdot\cdot\cdot (P_n, h_n)\}$, where $P_i$ is the set of pixels in the $i$-th component, and $h_i$ is its corresponding color in the HSV color histogram with $512$ bins. We use $|P_i|$ to denote the number of pixels in $P_i$. We extract the following features based on the above definition:
\begin{itemize}
\item $f_{10}= n$, which indicates the number of connected coherent components in the image.

\item $f_{11}=\frac{\max_i |P_i|}{|I|}$, which indicates the size of the largest component relative to the whole image.

\item $f_{12}= \frac{\max\limits_{j, j\neq \arg \max\limits_{i} |P_i|} |P_j|}{|I|}$, representing the size of the second largest connected coherent component relative to the whole image. 

\item $f_{13}=\mathrm{rank}(h_i), i=\arg \max_j |P_j|$, indicating the rank of the bin, considering the bin size in descending order, associated with the largest connected coherent component in the image. For example, the value of this feature is $1$ if the bin associated with the largest coherent component, $\arg \max_j |P_j|$, is the largest bin in the color histogram as well; $\max_i |h_i|$ where $h_i$ is the size of the $i-{th}$ bin in the color histogram. This feature indicates how the colors are disperse in the image. We expect to have value $f_{13}=1$ if the colors in the images are not very randomly distributed. It means the pixel with the same colors are mostly connected together.


\item $f_{14}=\mathrm{rank}(h_i), i=\arg \max_{j, j\neq \arg \max_k |P_k|} |P_j|$, similar to $f_{13}$, it shows the bin rank, considering the bin size in descending order, of the second largest connected coherent component in the image.
\end{itemize}

\subsubsection{Hue Distribution}
\label{ss:hue-dist}
In this section we introduce three features based on the hue in HSV color space. We quantize hues in an image in a similar way as in \cite{Li09} by eliminating the pixels with saturation and value less than $0.2$. This will eliminate all the pixels with white or black colors. Then we calculate the hue histogram of remaining pixels with $20$ different bins, $18^{\circ}$ for each bin, which results in $\mathcal{H}_{hue}=\{h_1, h_2,\cdot \cdot \cdot , h_{20}\}$ where $h_i$ indicates the set of pixels in $i$-th bin. We then extract the following features:
\vspace{-0.1in}
\begin{itemize}
\item $f_{15} = \sum_{i=1}^{20} \textbf{1}(|h_i| \geq c_5|I|)$ where $c_5=0.01$ in our experiments. This feature indicates the number of dominant hues in an image.

\item $f_{16}=\max_{i,j} \parallel |h_i|-|h_j|\parallel$ where $|h_i|\geq c_5|I|$, and $\|\cdot\|$ is the arc length. This feature indicates the largest contrast between two dominant hues in the image. 

\item $f_{17}=\mathrm{std}(\Phi)$ where $\Phi=\{\cup_{i\in I}\;  \parallel h_i(i)-0\parallel\}$ and $\parallel . \parallel$ is the arc length value. This feature indicates the standard deviation of all pixel's hues distance from the origin $0$. It simply can determine how much the hue colors in an images has been distributed from each other.
\end{itemize}

\subsubsection{Lightness Features}
\label{ss-lightness}
We use the lightness $L$ in the HSL color space to calculate feature $f_{18}$ and  $f_{19}$. In the HSL color space, $L$ value is small when the color is white and is large when the color is black. The $L$ value in HSL color space can be calculated as follows:
\begin{equation} 
L(x)= \frac{\max\left(r(x), g(x), b(x)\right) + \min\left(r(x), g(x), b(x)\right)}{2},
\end{equation}
where $r(x), g(x), b(x)$ denotes the R, G, B values of pixel $x$ in RGB color space. We calculate two lightness features as:

\begin{itemize}
 \item $f_{18}=\frac{1}{|I|} \sum_{x\in I} L(x)$, the average lightness of pixels in the image.
 \item $f_{19}=\mathrm{std}(L(\cdot))$, the standard deviation of lightness of all pixels in the image.
\end{itemize}

\subsection{Local Features}
Local features represent a set of features extracted from specific parts of the image rather than the whole image. We apply the normalized cut segmentation method \cite{Shi00} to partition the image into $5$ smaller segments. Let  $\mathcal{S}=\{S_1,S_2,\cdots,S_5\}$ indicate the set of $5$ different segments where $S_i$ is the set of pixels in segment $i$. Note that a segment is considered as noise and is dropped if it is smaller than $5\%$ of the image. We develop the following features based on the segmentation result.

\subsubsection{Segment Size}
\label{ss-s-Segmentats-size}
Two features are extracted from segment size as follows:

\begin{itemize}
	\item $f_{20} = \frac{\max_i |S_i|}{|I|}$, indicating the size of the largest segment relative to the whole image.
\item $f_{21} =\frac{1}{|I|}\max_{i,j}\big||S_i|-|S_j|\big|$, indicating the contrast among the segmentation sizes of the image. 
\end{itemize}

\subsubsection{Segment Hues}
\label{ss-s-segment-hues}
Similar to section \ref{ss:hue-dist}, we generate the hue histogram of each segment. We define the set of hue histograms of all $5$ segments as $\mathcal{H}^s_{hue}=\{h_{1,1},h_{1,2},\cdot \cdot \cdot, h_{1,20}, h_{2,1}, \cdot\cdot\cdot, h_{5,20}\}$ where $h_{i,j}$ indicates the set of pixels that fall in the $j$-th bin of $i$-th segment. Then we extract five features to capture different hue properties. Below we describe the formal definition of developed features:
\vspace{-0.1in}
\begin{itemize}
\item $f_{22}=\sum_{j=1}^{20} \textbf{1}(|h_{i,j}\geq c_6|I|)$ where $i=\arg\max_i S_i$ and $c_6=0.01$. This feature denotes the number of image-wide dominant hues in the largest segment. In general, we would like to have most of the image hues in the largest segment.

\item $f_{23}=\sum_{j=1}^{20} \textbf{1}(|h_{i,j}\geq c_6|S_i|)$ where $i=\arg\max_i S_i$. This feature denotes the number of segment-wide dominant hues in the largest segment.

\item $f_{24}=\max\limits_{i}q_i$ where $q_i = \sum_{j=1}^{20} \textbf{1}(|h_{i,j}\geq c_6|S_i|)$ is the number of dominant hues in $i-{th}$ segment. This feature essentially denotes the largest number of dominant hues in one segment. We would like to have the same value as $f_{23}$ for this feature illustrating that the largest segment has the largest number of dominant colors.

\item $f_{25}=\max\limits_{i,j}|q_i - q_j|$. This feature denotes the contrast of the number of dominant hues among the segments. We usually do not like to have lots of different hues in one segment and a few hues in another segment in an image. We expect to have unappealing images with large value for $f_{25}$.

\item $f_{26}= \max\limits_{j,k} \|h_{i,j}-h_{i,k}\|$ where   $|h_{i,j}|, |h_{i,k}|\geq c_6|S_i|$,  $i=\arg\max_i |S_i|$ and  $\|\cdot\|$ is the arc length distance. This feature captures the contrast of number of pixels among the hue bins in the largest segment. In general, we expect to have an appealing image with one bin dominating the largest segment in addition to a few more small bins. This makes the contrast value very large.

\item $f_{27}=\mathrm{std}(T(\cdot))$ where $T(i)=\max_{j,k}$, $|h_{i,j}|, |h_{i,k}|\geq c_6|S_i|$. This feature returns the standard deviation of  contrast  among the segments. If we have different hue contrasts among different segments, this feature will achieve a significant value.
\end{itemize}

\subsubsection{Segment Color Harmony}
\label{ss-s-color-harmony}
Two features are  extracted based on the largest segment color harmony. Feature  $f_{28}$ is the minimum deviation from the best fitted color harmony model for the largest segment, and feature $f_{29}$ is the average deviation of the best two fitted color harmony models for the largest segment. The details of color harmony models have been introduced in section \ref{ss-model-based-color-harmony}.

\subsubsection{Segment  Lightness}
\label{ss-s-lightness}
Three segment lightness features are extracted using similar method as in section ~\ref{ss-lightness}:
\vspace{-0.1in}
\begin{itemize}
\item $f_{30}$: average lightness in the largest segment.
\item $f_{31}$: standard deviation of average lightness among the segments.
\item $f_{32}$: contrast of average lightness among the segments.
\end{itemize}

\subsection{Advanced features}
In this section we develop a set of features based on more complicated algorithms. Most of the advanced features are based on the saliency map of the image which determines the visually salient areas in the image that are more likely to be noticed by the humans. We also extract two additional features related to the number of characters and number of faces in an image. Below we describe the details of these features.

\subsubsection{Saliency Features}
\begin{figure}
\begin{center}
\addtolength{\tabcolsep}{-0.7em}
\begin{tabular}{c c}
\includegraphics[width=1.0in,height=1.4in]{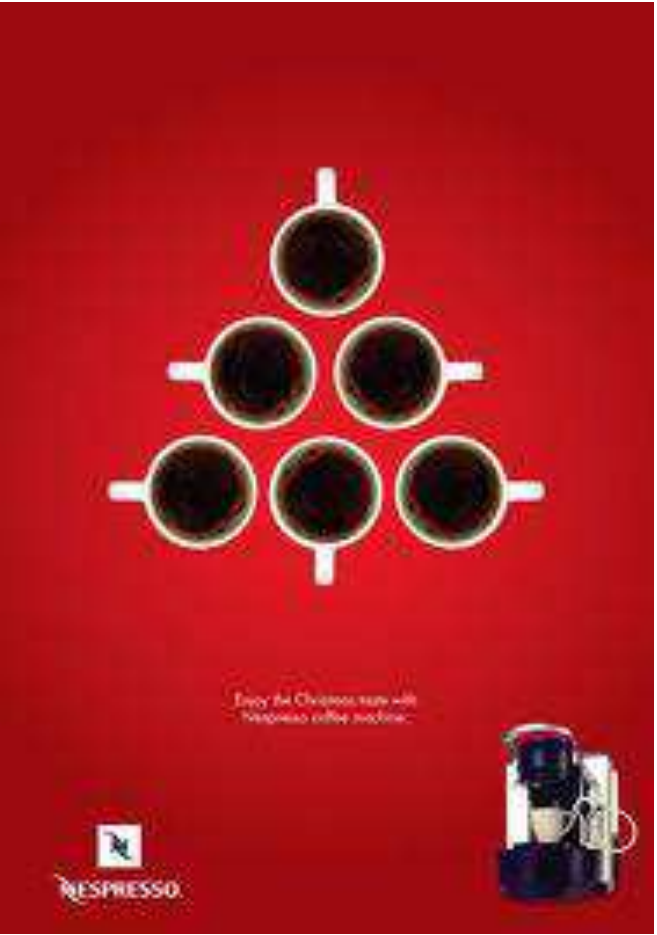} & \space \space \space
\includegraphics[width=1.0in,height=1.4in]{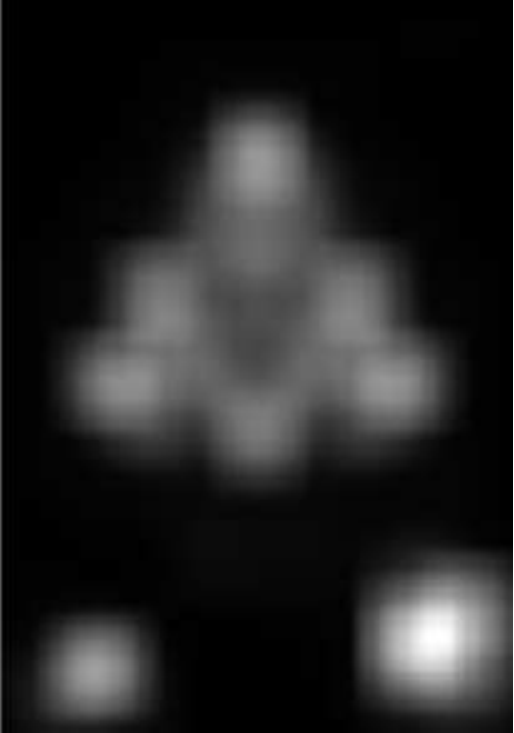}\\
\end{tabular}
\end{center}
\caption{The saliency map of an image. Left: original image. Right: saliency map.}
\label{fig:sal-example}
\end{figure}

Saliency computation is a well known phenomenon in human vision where attention tends to be drawn to interesting parts of an image that appear visually different from the rest of the image (e.g., a red coke can in a green background appears salient and is immediately noticed, while the same coke can in an orange-reddish background is not salient and less likely to be noticed).  We compute saliency according to the algorithm described in ~\cite{Hou07}. Figure~\ref{fig:sal-example} shows the saliency output of the algorithm presented in~\cite{Hou07} for a sample creative. The areas with higher lightness in the saliency map indicate more salient part of the image.

The saliency algorithm returns a matrix $\tau$ (also referred to as saliency map) where $\tau(i,j)$ represents the saliency value of  pixel $(i,j)$. We also extract a binary image based on the saliency map, by setting a threshold $\alpha$ to the saliency map where the pixels with saliency value larger than $\alpha$ are set to $1$ and the rest of the pixels are set to $0$. Similar to \cite{Hou07}, the parameter $\alpha$ is set as $\alpha=3\bar{\tau}$ where $\bar{\tau}=1/n\sum_{i,j}\tau(i,j)$ is the average saliency value in the image. After this binarization, we have some connected components with value 1. These components indicate  saliency areas, and the other parts of the image are considered as background. Then we extract the following features based on the saliency results, saliency map and binary saliency map.

\begin{itemize}
\item $f_{33}$: background size. Salient objects usually appear in the foreground and not in the background. Therefore we return the size of the background as a function of image size which is calculated as:
$f_{33}=\frac{\sum_{i,j} \textbf{1}(\tau(i,j)<\alpha)}{|I|}.$

\item $f_{34}$: number of connected components in the binary map.

\item $f_{35}$: size of the largest components in the binary saliency map relative to the whole image.

\item $f_{36}$: average saliency value of the largest component in the binary saliency map.

\item $f_{37}$: number of connected components in the image background. In some images, the saliency areas can divide the background into several disconnected segments. Usually it is not desirable to have multiple background components.

\item $f_{38}$: size of the largest connected component in the background relative to the whole image. If the number of connected components in the background is equal to one, then this feature has the same value as $f_{33}$.

\item $f_{39}$: distance between connected components. Let the set $\mathcal{C}=\{c_1,c_2,\cdots, c_n\}$, $c_i=(x_i,y_i)$ indicates the set of $n$ different points such that each  $c_i$ indicates a pixel corresponding to the center of mass of the $i$-th saliency area. To make the rest of the computation scale independent from the image size, we update the properties of each point $c_i$ as $s_i=(x_i/I_x, y_i/I_y)$ such that $I_x$ and $I_y$ are the horizontal and vertical size of the image. Then we build up a complete weighted graph given the set $\mathcal{C}$ such that the weight $w_{i,j}$ between two vertices $c_i, c_j$ is calculated as $w_{i,j}=\| s_i-s_j\|_2$. Then we return the summation of all edge weights as the distance between connected components.

\item $f_{40}$: distance from the rule of third points. Professional photographers usually locate their main object in one of the four interest points based on the rule of third. The four interest points in rule of third is the intersection of two vertical and two horizontal lines dividing the image into $9$ equal segments. Figure~\ref{fig:rule-of-third} shows the four interested points based on rule of third. This is an important feature in photo beauty evaluation~\cite{Li09}, motivating us to investigate its effect in creative performance. We define this feature as the minimum distance from the center of mass of the largest saliency area  to one of the four interest points based on rule of third.

\begin{figure}
\begin{center}
\addtolength{\tabcolsep}{-0.7em}
\begin{tabular}{c }
\includegraphics[width=1.5in,height=1.2in]{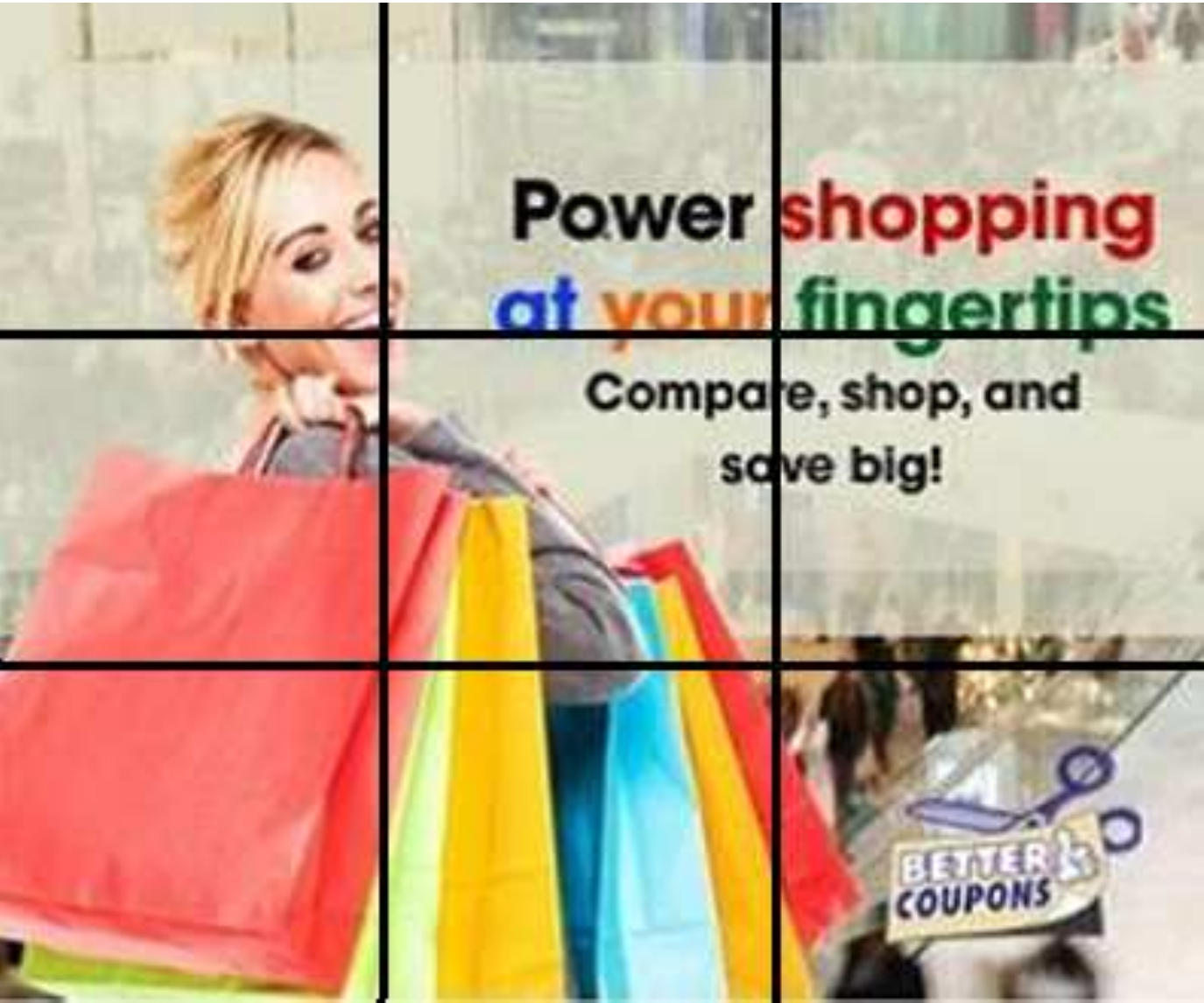}\\
\end{tabular}
\end{center}
\caption{The four interested points based on rule of third.} \vspace{-0.1in}
\label{fig:rule-of-third}
\end{figure}

\item $f_{41}$: distance from the center of image. This feature is the distance of saliency components to the center of image which is the most focused part of an image. The overall distance from the centers of all connected components to the center of image is returned as feature $f_{41}$. Note that for both features $f_{40}$ and $f_{41}$, we normalize the position of each pixel similar as feature $f_{39}$. 
\end{itemize}

\subsubsection{Number of Characters}
We consider the number of characters in an image as feature $f_{42}$. We tried a number of OCR toolbox and one of them provides us with appropriate results considering the number of characters in ads\cite{OCR}. Note that we are interested in the number of characters in the image regardless of its meaning. To evaluate the accuracy of the OCR toolbox, we counted the true number of characters in $100$ random images and compared it to the returned number of characters from the OCR toolbox. We found strong linear correlation of $0.80$, suggesting that our toolbox is reasonably accurate in evaluating the number of characters in images. Note that extracting the exact text from ad creatives is challenging as they often appear in different fonts, sizes and orientations.

\subsubsection{Number of Faces}
The last feature, $f_{43}$, captures the effect of the human face appearance on creative performance. In \cite{Isola2011} the authors concluded that the human appearance in an image could make the image more memorable. This motivates us to test whether face appearance affects creative performance. We count the number of faces in an image using an available toolbox \cite{FaceDet}. Our toolbox is reasonably accurate and has a correlation more than $0.9$ with the true number of faces in images in our experiments with a sample size of $100$.

\section{Experimental Results}
\label{sec:experimental-results}
In this section we present the algorithms and experiments we designed to evaluate the relationship between visual features and the performance of creatives in online display advertising.

\subsection{Data Set}
We extracted creatives of advertising campaigns from the world's largest online advertising exchange system, RightMedia. We filtered out animated creatives because our features are designed for static images. We also calculated the average CTR of these creatives from online serving history log during a two-month period.

\begin{figure}
\begin{center}
\subfigure[Data set ID2]{\label{fig:id2}\includegraphics[width=0.240\textwidth]{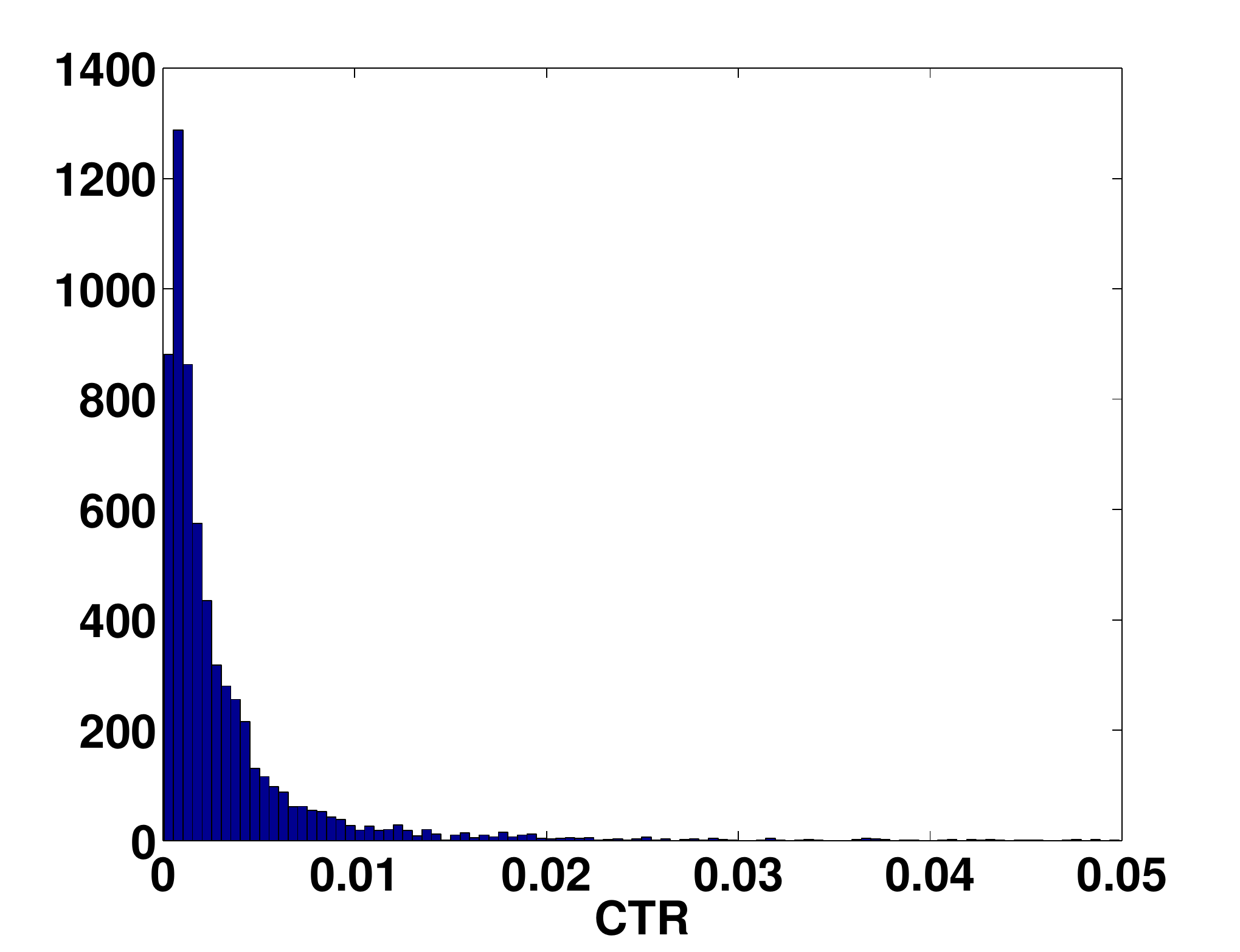}}
\hspace{-0.03\textwidth}
\subfigure[Data set ID6]{\label{fig:id6}\includegraphics[width=0.240\textwidth]{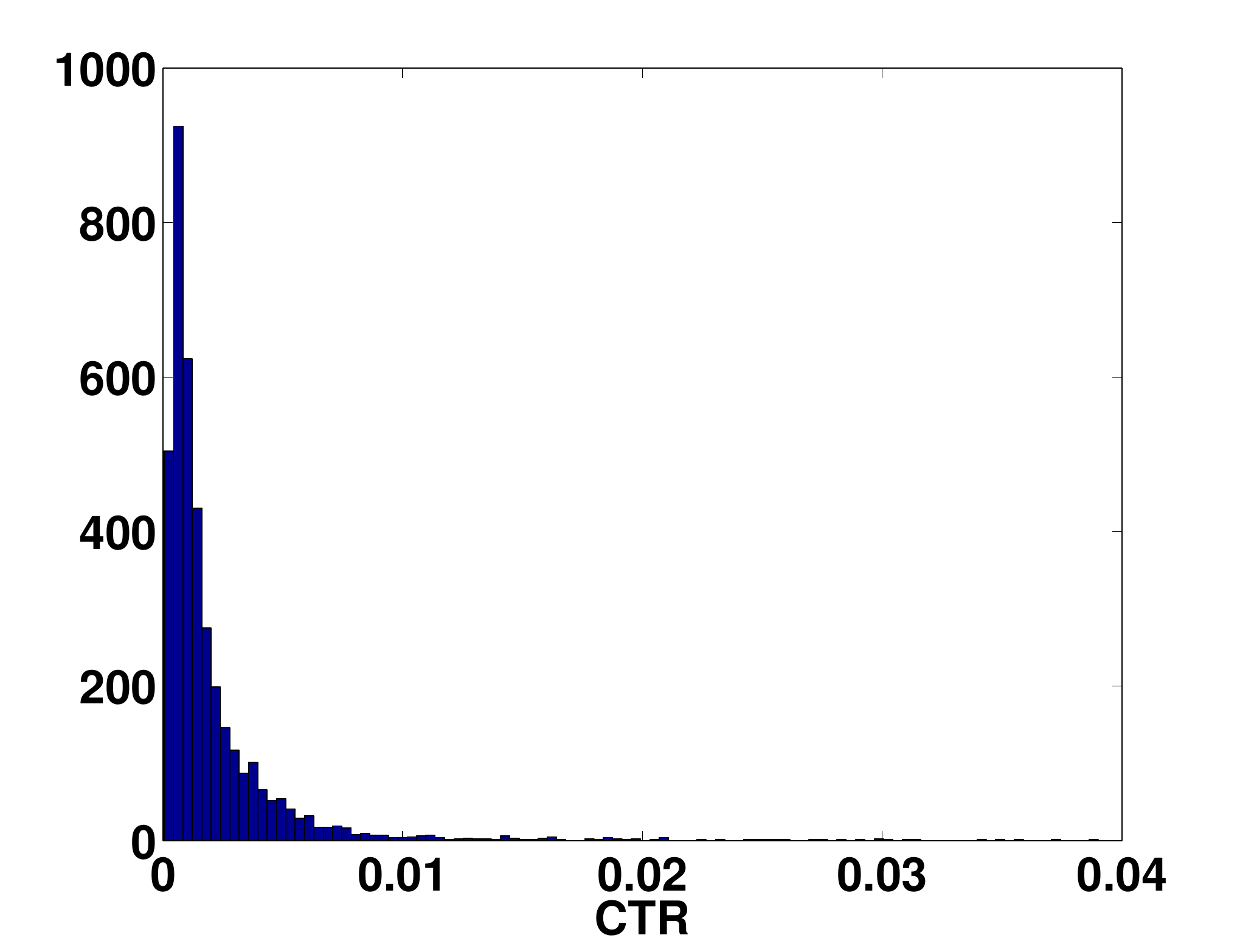}}
\caption{The CTR distribution of two data sets. }
\label{fig:ctr-dist}\vspace{-0.15in}
\end{center}
\end{figure}

As discussed in Section~\ref{sec:introduction}, the performance of creatives is determined by many factors. One important factor is the ad position in the webpage. Generally the available position of a creative on a webpage is determined by the creative's size. To remove the impact on performance introduced by ad position (and size), we create two different data sets, each of which consists of creatives with the same size. The first data set, ID$2$, consists of $6272$ creatives with size $250\times300$ pixels, and the second data set, ID$6$, includes $3888$ images with  $90\times730$ pixels. All of the creatives have a minimum of $100K$ impressions guaranteeing that their CTRs have converged to their true values. The CTR distribution of each data set is shown in Figure \ref{fig:ctr-dist}.

We further created two sub-categories from data set ID$2$: ``dating'' with $927$ images and ``traveling'' with $599$ images. Since there are not many images in these two categories, we consider the images with a minimum of $20k$ and $10k$ impressions for ``dating'' and ``traveling'' respectively.

\subsection{Learning Methods}
The main goal of this work is to study the relationship between the performance of creatives and their visual features. In the first step we try to predict CTR from  visual features using regression methods. We used three different regression algorithms to predict CTR, 1) Linear Regression (LR), 2) Support Vector Regression with RBF kernel(SVR), and 3) Constrained Lasso (C-Lasso) which is a modification to Lasso~\cite{tibshirani96}.

We used LIBSVM \cite{libsvm} to implement the SVR  and performed cross validation to determine the parameters of the model. We describe our constrained Lasso optimization approach as follows. Suppose we have a set of $n$ creatives at disposal and the visual features of these creatives are represented as a matrix $A\in \textbf{R}^{d\times n}$ such that $A=(\mathbf{a}_1,\mathbf{a}_2,\cdots, \mathbf{a}_n)$ where $\mathbf{a}_{k}\in\textbf{R}^d$ is a column vector representing the $d$ dimensional visual features of creative $k$. In our experiment $d=43$. The CTR values of the $n$ creatives are represented as a vector $\mathbf{y}=({y_1,\cdots,y_n})^{\top}\in\textbf{R}^n$ where each $y_k$ is the CTR of the $k$-th creative. We bound the CTR of each creative by $y_{min}\leq y_i \leq y_{\max}$ where $y_{min}$ and $y_{max}$ can be obtained from online serving history log. To predict CTR of the creatives, we try to solve the following optimization problem:
\begin{equation}
\begin{aligned}
\min_w & & \| A^{\top}\mathbf{w}-\mathbf{y} \|_F^2+ \lambda \| \mathbf{w} \|_1 \\
s.t. & & y_{min} \leq  A^{\top} \mathbf{w}\leq y_{max}
\end{aligned}
\end{equation}
where $\|\cdot\|_F^2$ is Frobinius-2 norm and $\|\cdot\|_1$ is $\ell_1$ norm, also called lasso. We call the above optimization problem as constrained Lasso (C-Lasso) and we used \cite{cvx} to find the solution of this optimization problem. Note that the proposed C-Lasso approach performs better than Lasso in our application.

\subsection{Evaluation}
In this section we present different evaluation methods to analyze the efficacy of the developed visual features in predicting the performance of creatives.

\subsubsection{CTR Prediction}
\begin{figure*}
\begin{center}
\begin{tabular}{c c c c }
\includegraphics[width=1.5in, height=1.4in] {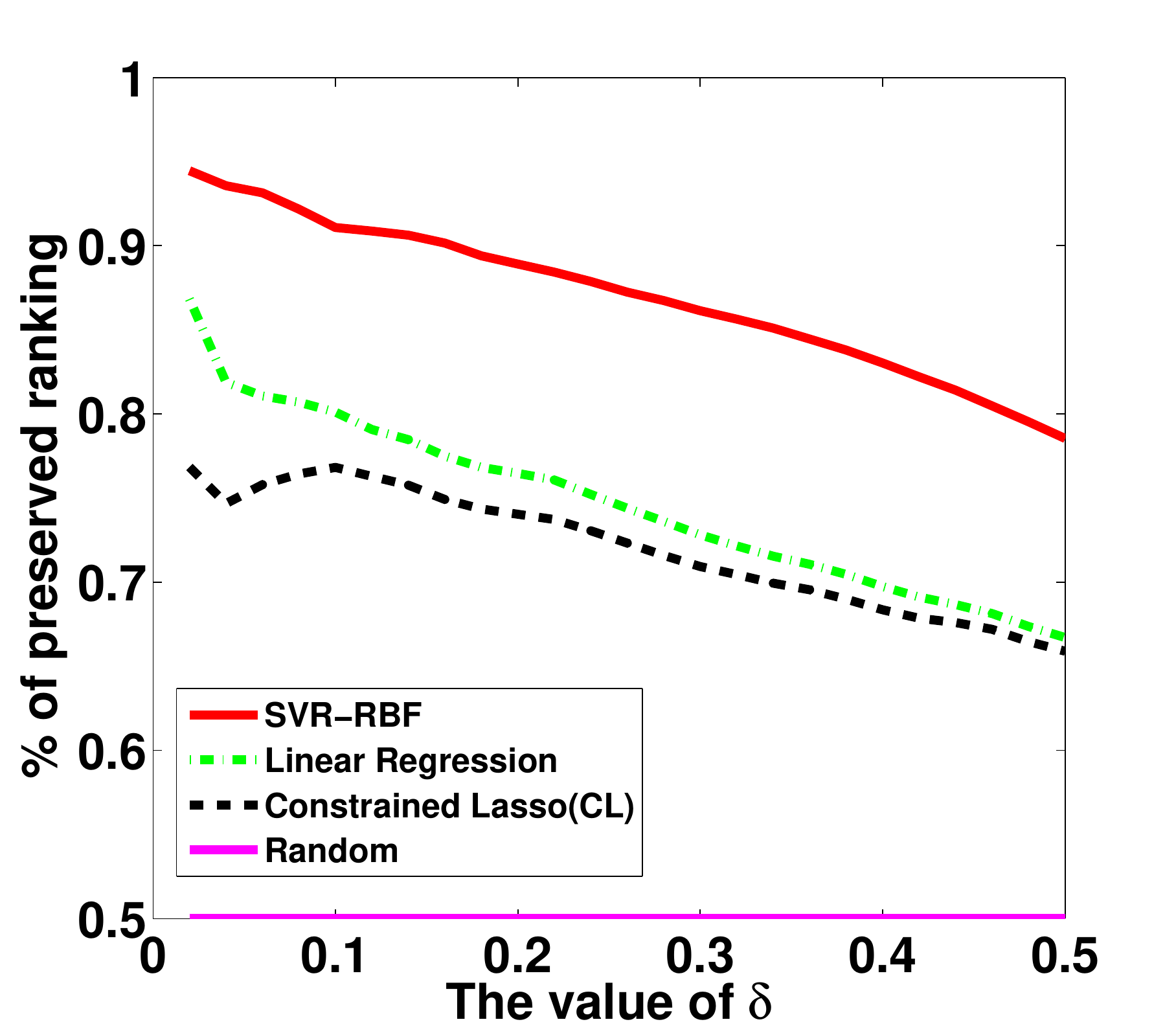} &
\includegraphics[width=1.5in, height=1.4in] {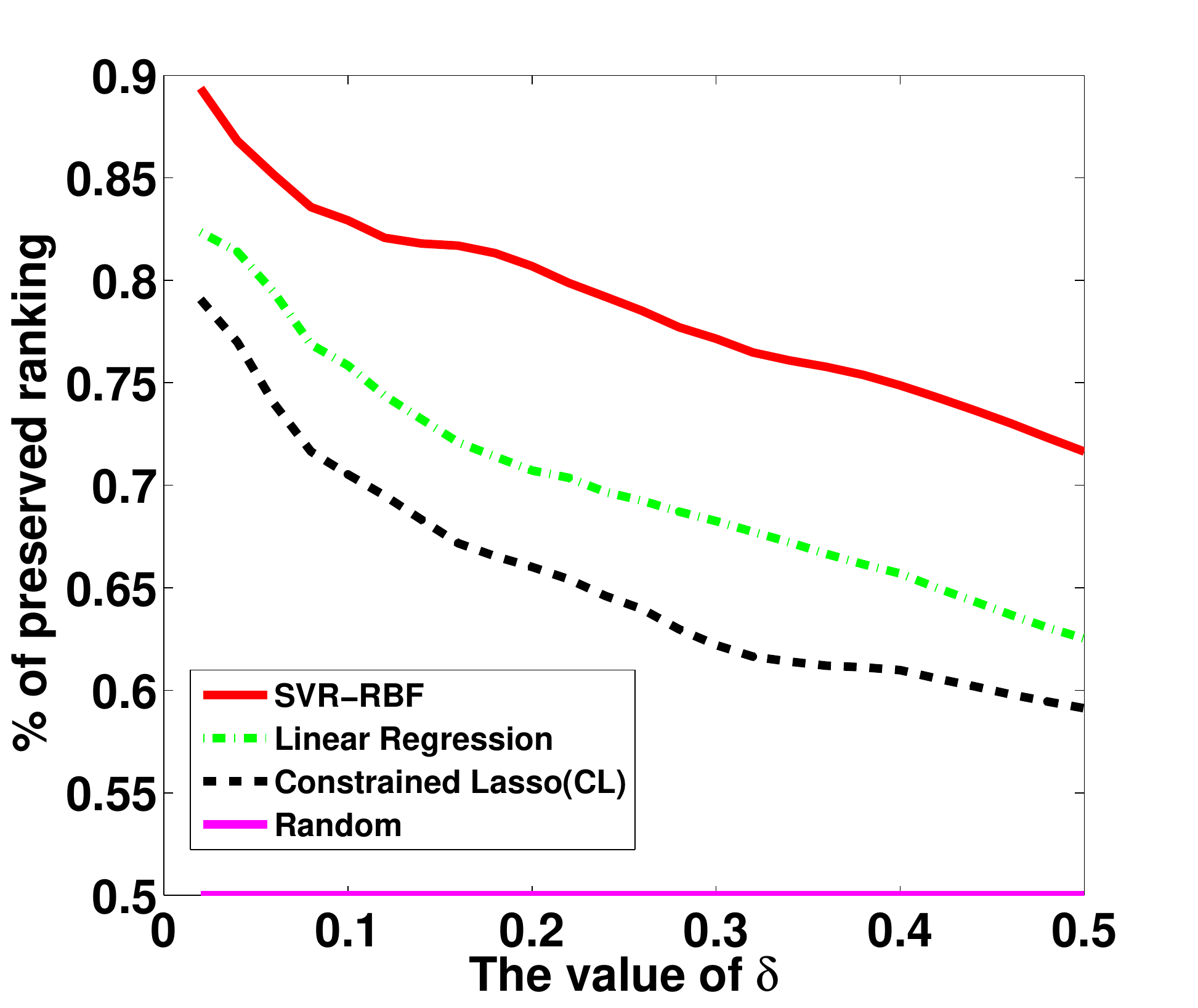} &
\includegraphics[width=1.5in, height=1.4in] {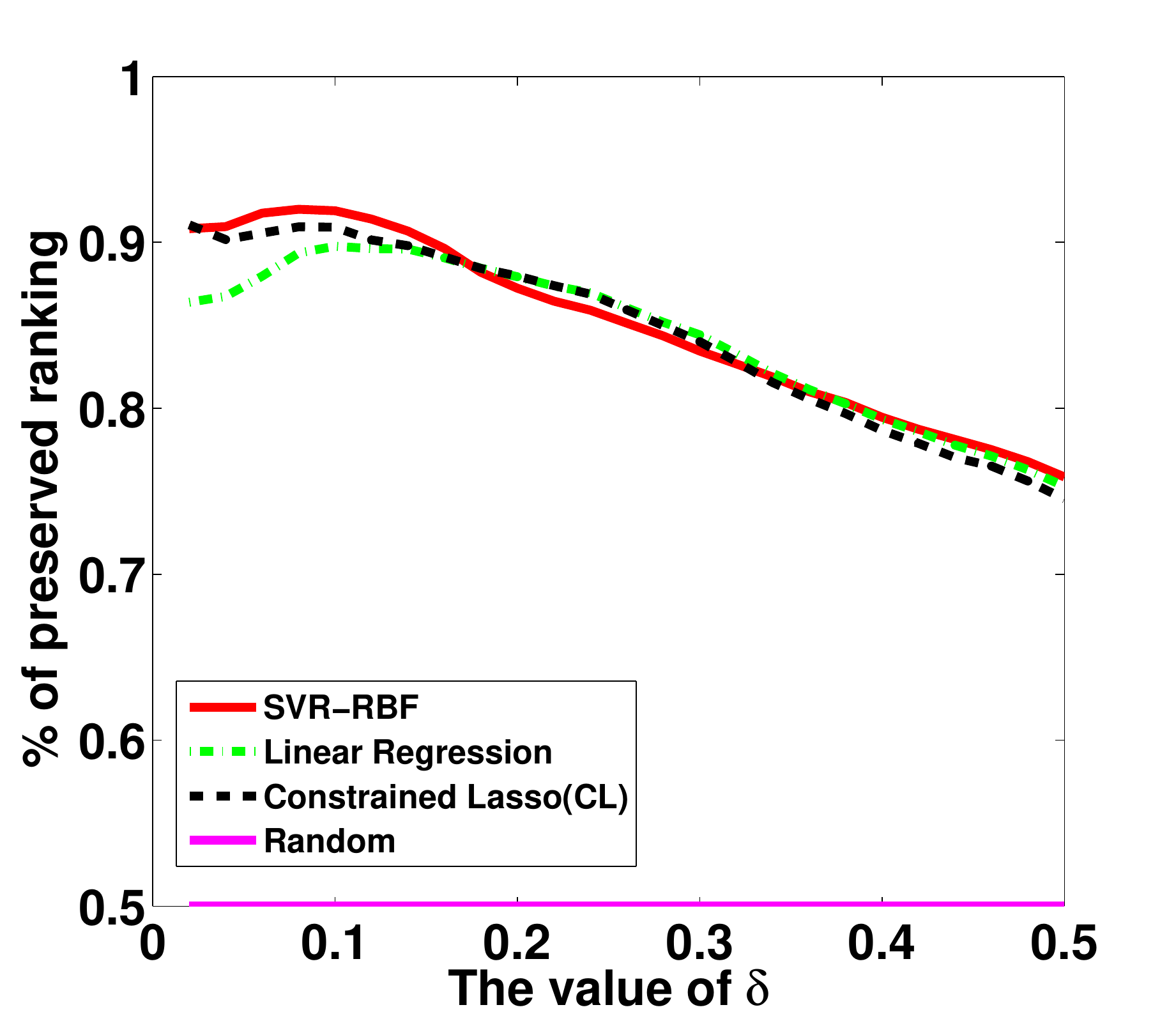} &
\includegraphics[width=1.5in, height=1.4in] {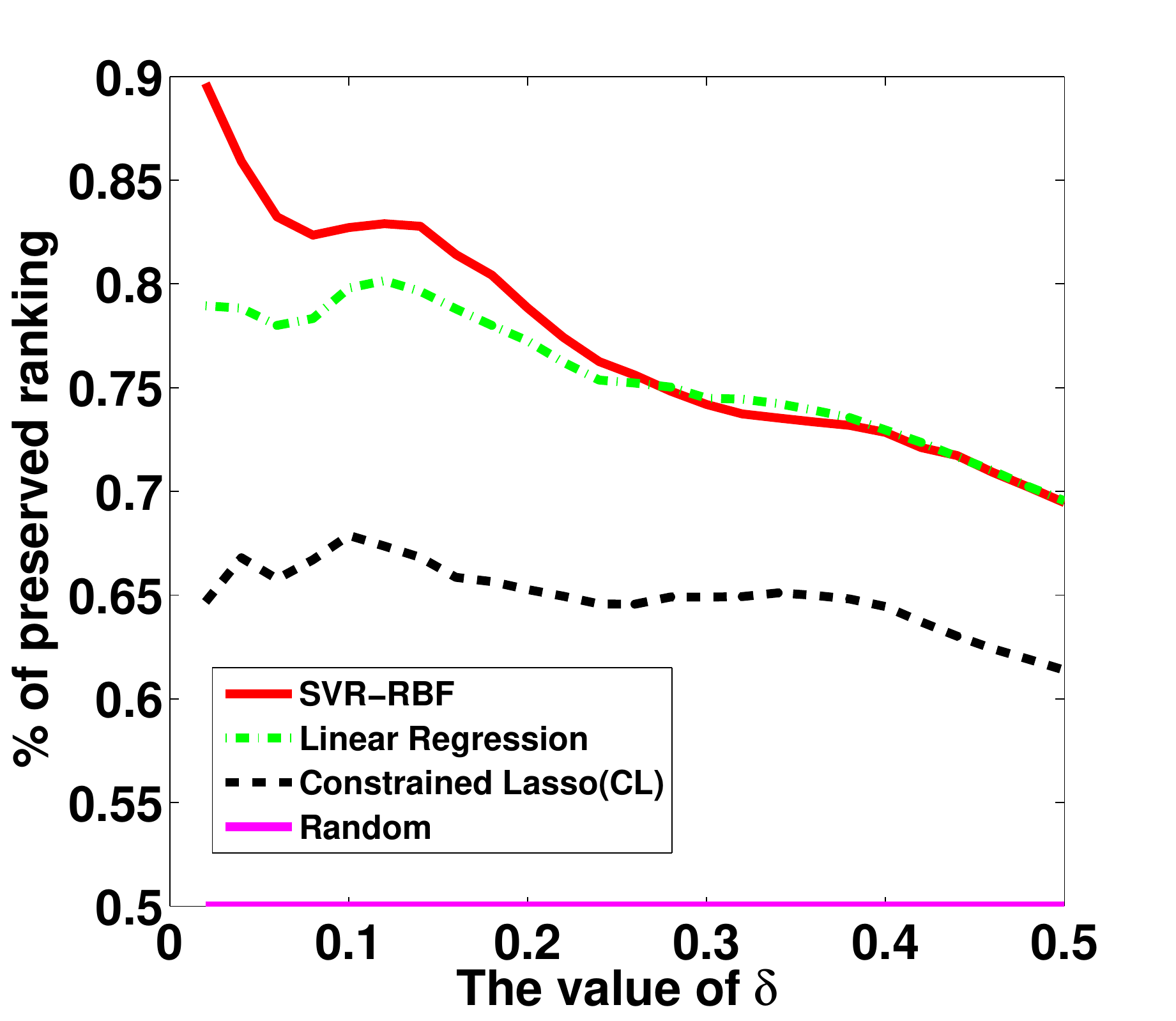}\\
ID$2$ &    ID$6$ &ID$2$-Dating & ID$2$-Traveling, IMP$\geq 10k$\\
\end{tabular}
\caption{The amount of preserved ranking for each method.}
\label{fiq:rank_100k}
\end{center}
\end{figure*}

To evaluate the CTR prediction accuracy of the algorithms, we run each algorithm for $200$ independent runs where in each run $80\%$ of each data set is selected randomly for training  and $20\%$ for testing. The accuracy evaluation results are reported over the prediction of the test data. Mean Squared Error (MSE) is used to measure the prediction accuracy for each algorithm as follows:
\begin{equation}
MSE=\frac{1}{n}\sum_{k=1}^{n} | y_k- \hat{y}_k|^2
\end{equation}
where $n$ is the number of test samples, $y_k$ is the true CTR of the $k$-th creative calculated from history log, and $\hat{y}_k$ is the predicted CTR.
To meaningfully interpret the MSE value, we introduce two baseline approaches, \emph{Random}  and \emph{Constant Mean}(CM) policy.

\begin{table}
\caption{The prediction accuracy of each method against \emph{Random} policy.}
\label{table:MSE-log-ctr}
\begin{center}
\begin{small}
\begin{tabular}{c c c c c c}
\textbf{Data set}& \textbf{Samples} & \textbf{CM} & \textbf{LR} & \textbf{C-Lasso} & \textbf{SVR} \\ \hline
ID$2$ &$6272$ &$1.71$ &$2.28$ &$2.22$ &$\textbf{3.27}$\\
ID$6$ &$3888$ &$1.75$ &$2.27$ &$2.14$ &$\textbf{2.77}$\\
ID$2$-Dating &$927$& $1.79$&    $2.65$ &$2.58$ &$\textbf{2.79}$\\
ID$2$-Traveling&    $599$ & $1.68$&    $2.13$ &$2.03$ &$\textbf{2.26}$ \\
\end{tabular}
\end{small}
\end{center}
\end{table}

The \emph{Random} policy simply samples from the CTR distribution of the training data to predict the CTR of each testing creative, while the CM policy assigns a constant value, $c_m$, to all ads where $c_m$ is the mean CTR of the training data.  Table \ref{table:MSE-log-ctr} shows the average results over $200$ independent runs for each algorithm. Each entry is the MSE value of the random policy divided by MSE value of each algorithm. Results show that we can perform up to $3.27$ times better than \emph{Random} policy in predicting the CTR from visual features only. All learners perform consistently better than baseline CM as well. This result demonstrates the non-trivial impact of visual appearance of the creative on its advertising performance.

\subsubsection{CTR Ranking}
We introduce a ranking criterion to investigate the ability of using visual features to rank the creatives by their CTRs. Given a test set of creatives, suppose $c_1^-,c_2^-,\cdots, c_k^-$ represent the $k$  images with the lowest CTR values and $c_1^+,c_2^+,\cdots, c_k^+$ represent the $k$ images with the highest CTR. Therefore we have $k^2$ pairs $(c_i^-,c_j^+)$ such that $ctr(c_i^-) \leq ctr(c_j^+)$ for $i,j\in\{1,\cdots,k\}$. We wish to know whether our prediction of CTR using visual features preserves the ranking of  pairs $(c_i^-,c_j^+)$. To test this, we change the value of $k$ as a function of test data size. We then measure the percentage of match between the predicted ranking of creatives, and the truly observed ranking in the test data.  The results over $200$ independent runs are shown in figure~\ref{fiq:rank_100k} for different data sets. The $x-$axis indicates the value of $\delta$ such that $k=\delta n$ for $\delta=0.02,0.04,0.06, \cdots , 0.50 $ where $n$ is the number of creatives in the data set, and $y-$axis represents the percentage of correctly ranked pairs.

Results show that SVR consistently outperforms other learners. As we increase the size of $k$, the percentage of correctly ranked predictions decreases for all learning algorithms. This is as expected, since differentiating the images of creatives which have CTRs close to the mean of the CTR distribution, using visual features only, is very difficult even for a human. Interestingly, the results show that by just using visual features, we can preserve more than $90\%$ of the ranking for data set ID$2$ (for $\delta=0.1$). This number remains high at $75\%$ when we consider all top-half images against low-half images for all data sets ($\delta=0.5$). This is an encouraging result that demonstrates the utility of visual features in predicting the ranking of CTR.

\subsubsection{CTR Classification}
\begin{figure}
\begin{center}
\begin{tabular}{c}
\includegraphics[width=1.7in, height=1.5in] {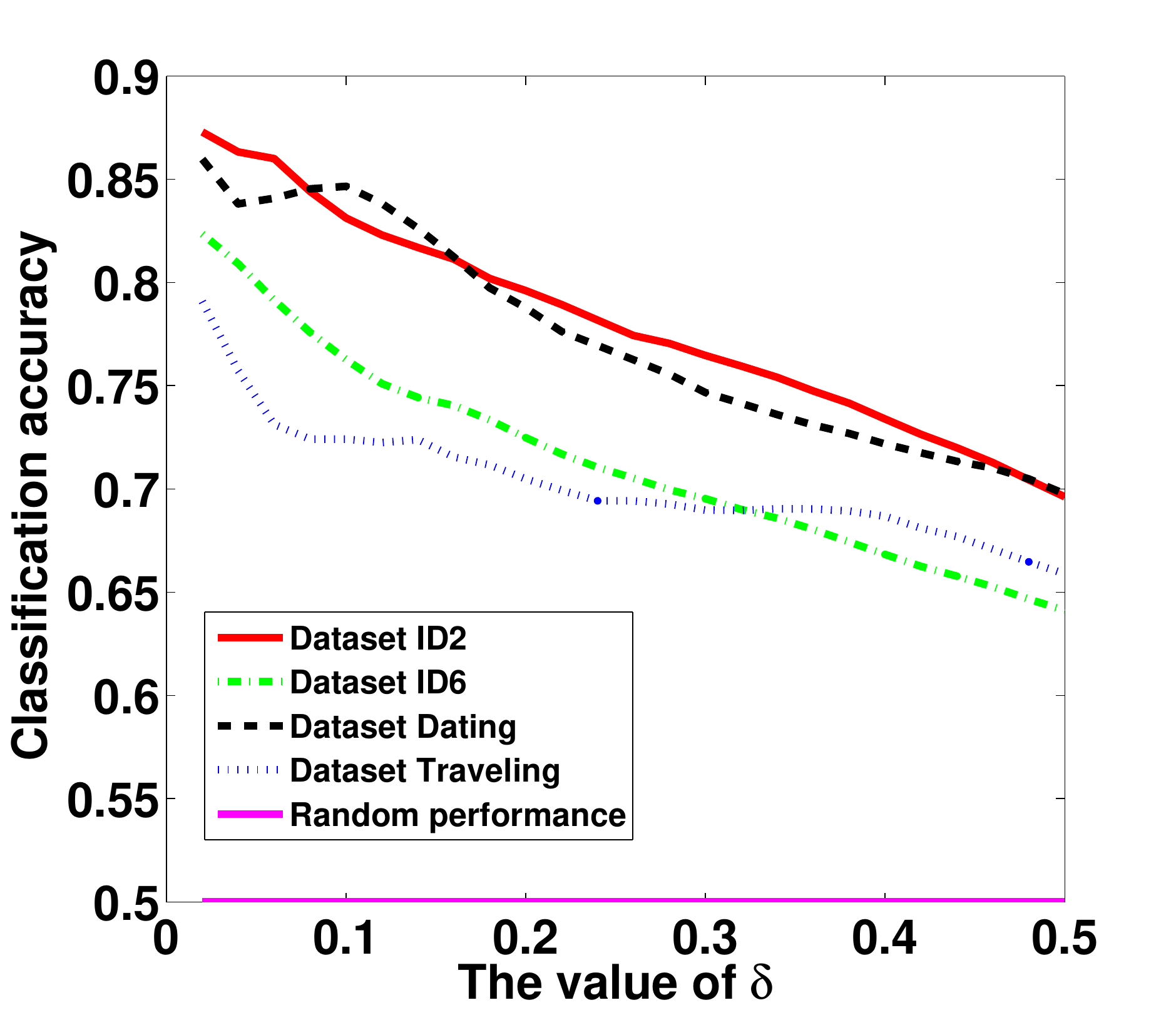}
\vspace{-0.1in}
\end{tabular}
\end{center}
\caption{The classification accuracy for each data set.}
\label{fiq:class}
\vspace{-0.1in}
\end{figure}

Previous studies in beauty evaluation \cite{Datta06,Ke06, Li09} mostly try to classify the images into high and low quality category rather than assigning scores to their beauty based on visual features. Similarly, we evaluate the performance of classifying the creatives into high (\textbf{+1}) and low (\textbf{-1}) CTR  category using visual features only. We use support vector machine with RBF kernel as our classifier. Similar to the previous section, we randomly separate $80\%$ of data as training and use the rest as testing data. Then, we train our classifier on creatives that belong to the top and bottom $30\%$ in CTR. In fact, we are disregarding $40\%$ of data that are close to the training data CTR mean, $\mu_{t}$, to reduce the noise for the classifier. Similar to the ranking experiments, we filter our test set by focusing on the $k$ creatives with highest CTR values (labeled as positive) and the $k$ creatives with the lowest CTR values (labeled as negative), where $k=\delta n$ is varied by changing $\delta$. We obtain the classification accuracy by comparing the predicted classes to the true classes obtained from real CTR values. Figure~ \ref{fiq:class} demonstrates the average classification accuracy over $200$ independent runs where each run uses randomly selected training and testing data. The $x$-axis indicates the value of $\delta$ and $y$-axis represents the classification accuracy for each data set given a fixed value of $\delta$. As seen in the figure, using visual features yields a classification accuracy of $70\%$ when $\delta=0.5$. Together with the previous results on predicting and ranking CTR, these results show the efficacy of using visual features of creatives in predicting CTR.
\vspace{-0.05in}

\subsection{Feature Selection}
The above analysis shows that visual features are useful in predicting the performance of creatives in online advertising. A natural question is to identify the visual features that have strong impact on ad performance. Such information could be very useful in many areas. For example, human graphic designers may use this information to guide their design of high-performance creatives. Smart ads system may use this information to dynamically generate creatives that are more appealing to online users. Ad exchange system may use this information to determine which creative will win in the auction marketplace for each advertising opportunity. In this section we conduct a series of experiments to select such important visual features.

We first calculate the Linear Correlation (LC) and Mutual Information (MI) between all features and CTR in each data set. Mutual information can provide us with the information of non-linear correlation between features. Note that, to calculate the mutual information between any pair of features $(X,Y)$, we discretized each feature and CTR values into $50$ equal intervals.
The results are shown in table \ref{table:fs-results}. The top $5$ features in each data set with highest absolute values are highlighted in bold. The table shows that there is no feature with high linear correlation or mutual information except  $f_{12}$ in data set ID$6$. Thus we use Forward Feature Selection (FFS) to select the top $k$ features.

Before running FFS, we first cluster the features based on the Normalized Mutual Information (NMI) of all feature pairs. We discretize each feature into $50$ equal intervals, and calculate NMI as follows:
\begin{equation}
NMI(X;Y)=\frac{I(X;Y)}{\sqrt{H(X)H(Y)}},
\end{equation}
where $H(X)$ is the entropy of random variable $X$. Then we cluster the features using the average linkage algorithm  \cite{Fionn83}. Two clusters are merged into one if their average NMI is at least $0.2$. This results in $20$ clusters for data set ID$2$ and $21$ clusters for data set ID$6$. The resulting clusters are shown in Table~\ref{table:fs-results}. In the table, $S_i$ represents a set of features in cluster $i$. We now apply a simple change to the FFS algorithm to select the top $k$ clusters rather than features. After selecting a feature by FFS, all the correlated features that belong to the same cluster are removed from the next steps of FFS. The selected top $k=10$ clusters are shown in table~\ref{table:selected-clusters}. Note that clustering the features in the above manner helps select different features (or feature sets) that are less correlated with each other. For example, all color harmony features are in the same cluster $S_4$. Therefore by selecting one of the features from this cluster, we indicate the importance of color harmony in CTR, and by removing the highly correlated features at each step in FFS, we can guarantee to select a set of features which are less correlated with each other. Below we investigate some of the selected clusters that are common to both data sets.

\begin{table}
\caption{The top 10 selected clusters by FFS.}
\label{table:selected-clusters}
\begin{center}
\begin{small}
\begin{tabular}{cc }
\textbf{Data Set}& \textbf{Selected Clusters}\\ \hline
ID$2$ &    $S_{5}$, $S_1$, $S_{19}$, $S_{17}$, $S_{13}$, $S_{18}$, $S_{20}$, $S_{10}$, $S_{11}$, $S_9$\\
ID$6$ &$S_1$, $S_2$, $S_{20}$, $S_5$, $S_{17}$, $S_{14}$, $S_9$, $S_{13}$, $S_4$, $S_{18}$\\
\end{tabular}
\end{small}
\end{center}
\end{table}

\begin{figure}
\begin{center}
\begin{tabular}{c c }
\includegraphics[width=1.5in, height=1.4in] {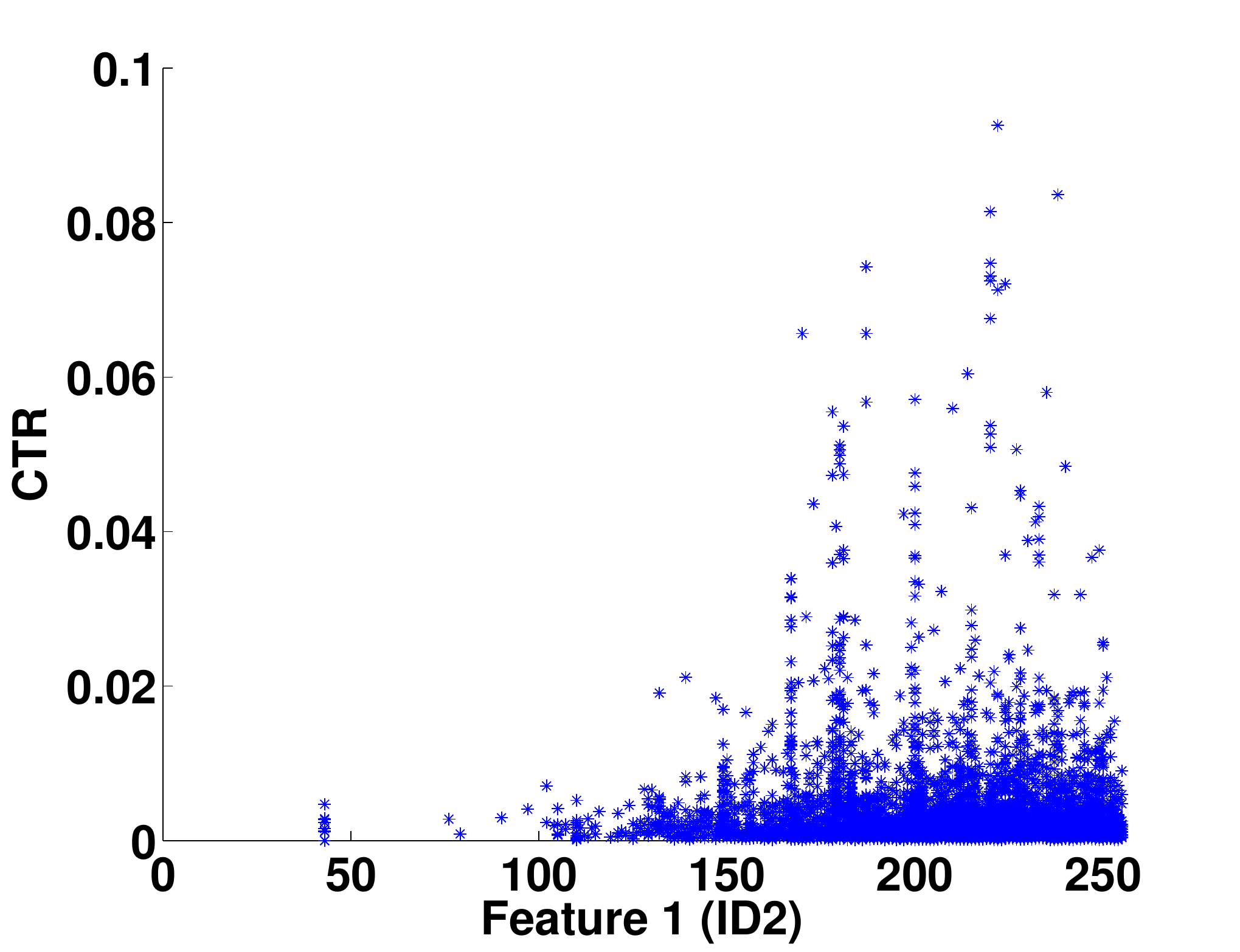} &
\includegraphics[width=1.5in, height=1.4in] {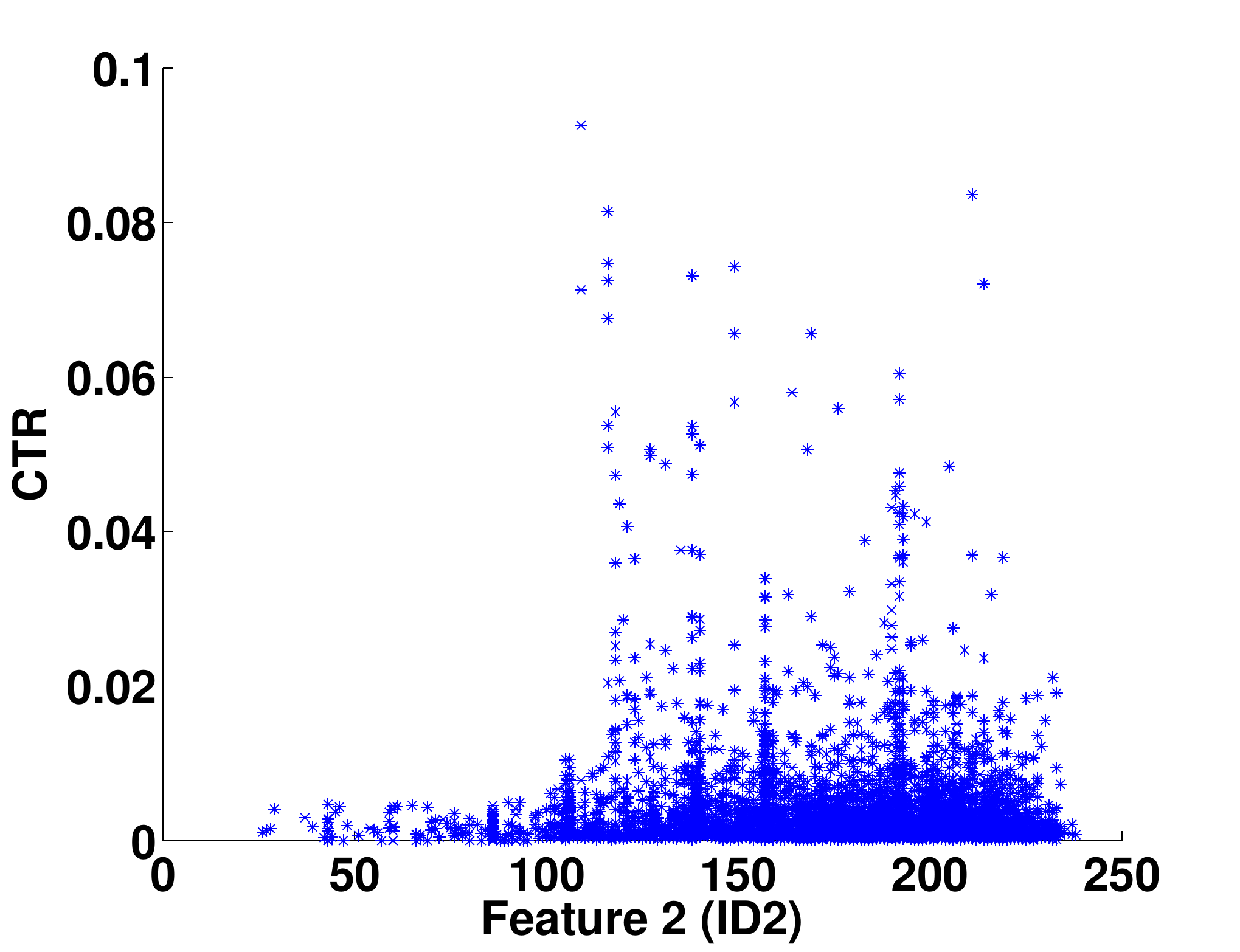}  \\
\end{tabular}
\caption{The scatter plot of $f_1$ and $f_2$ against CTR.}
\label{fig:s1-dis}
\end{center}
\end{figure}

Table \ref{table:selected-clusters} shows that $S_1$ is the best feature set (or cluster) for data set ID$6$ and the second best set for data set ID$2$ which illustrates the importance of set $S_1$. $S_1$ consists of the gray level features $f_1$ and $f_2$ of the image. 
The scatter plot of both features in data set ID$2$ is shown in Figure\ref{fig:s1-dis} (the scatter plot in data set ID$6$ is similar). Figure \ref{fig:s1-dis} shows that for creatives with small value in both features, high CTR value is unlikely, and creatives with high CTR values should have high values in these two features. This is consistent with the intuition that creatives with higher contrast should perform better. Note that having high values in these two features does not guarantee a high CTR value.

\begin{figure}
\begin{center}
\begin{tabular}{c c }
\includegraphics[width=1.5in, height=1.4in] {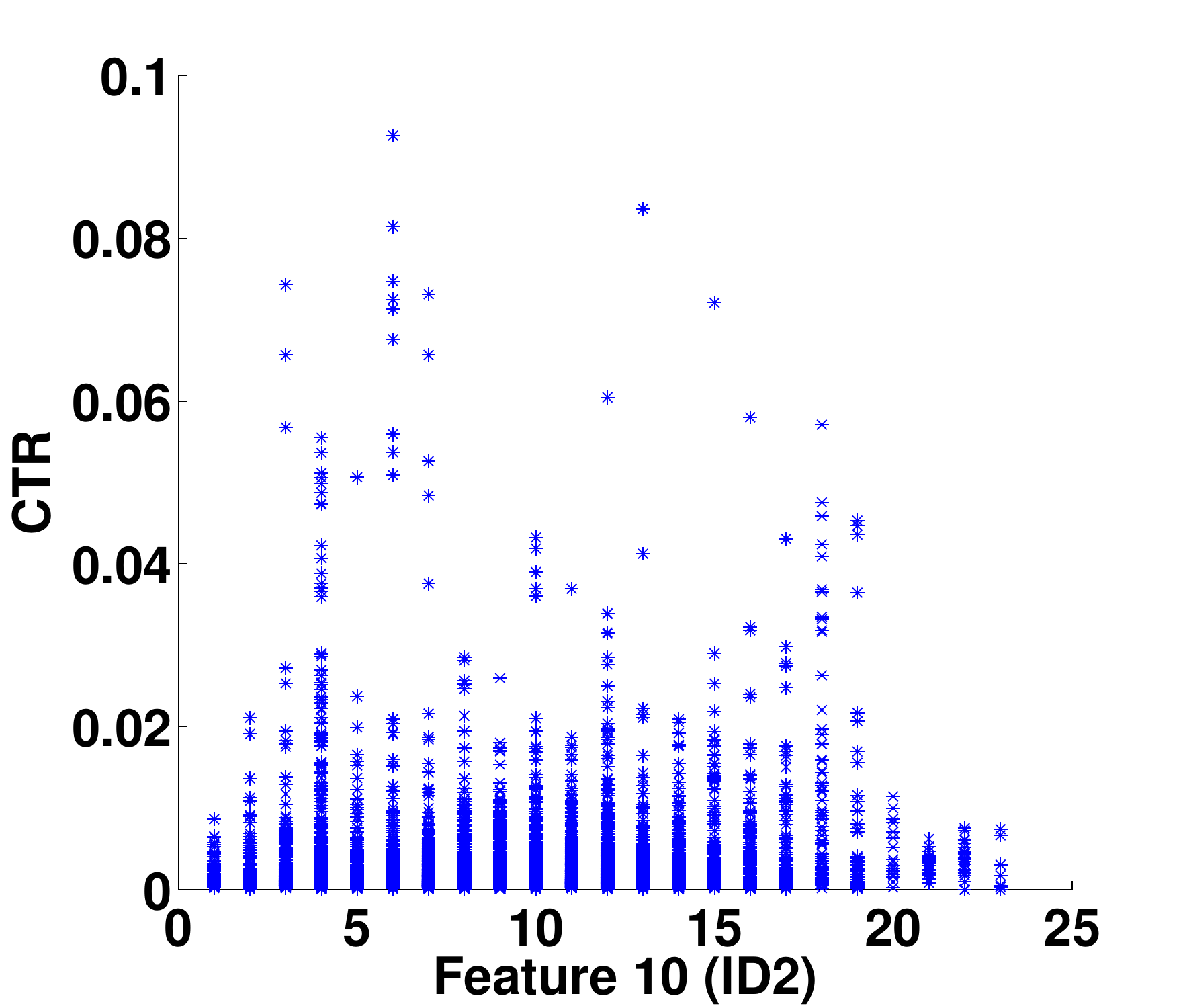} &
\includegraphics[width=1.5in, height=1.4in] {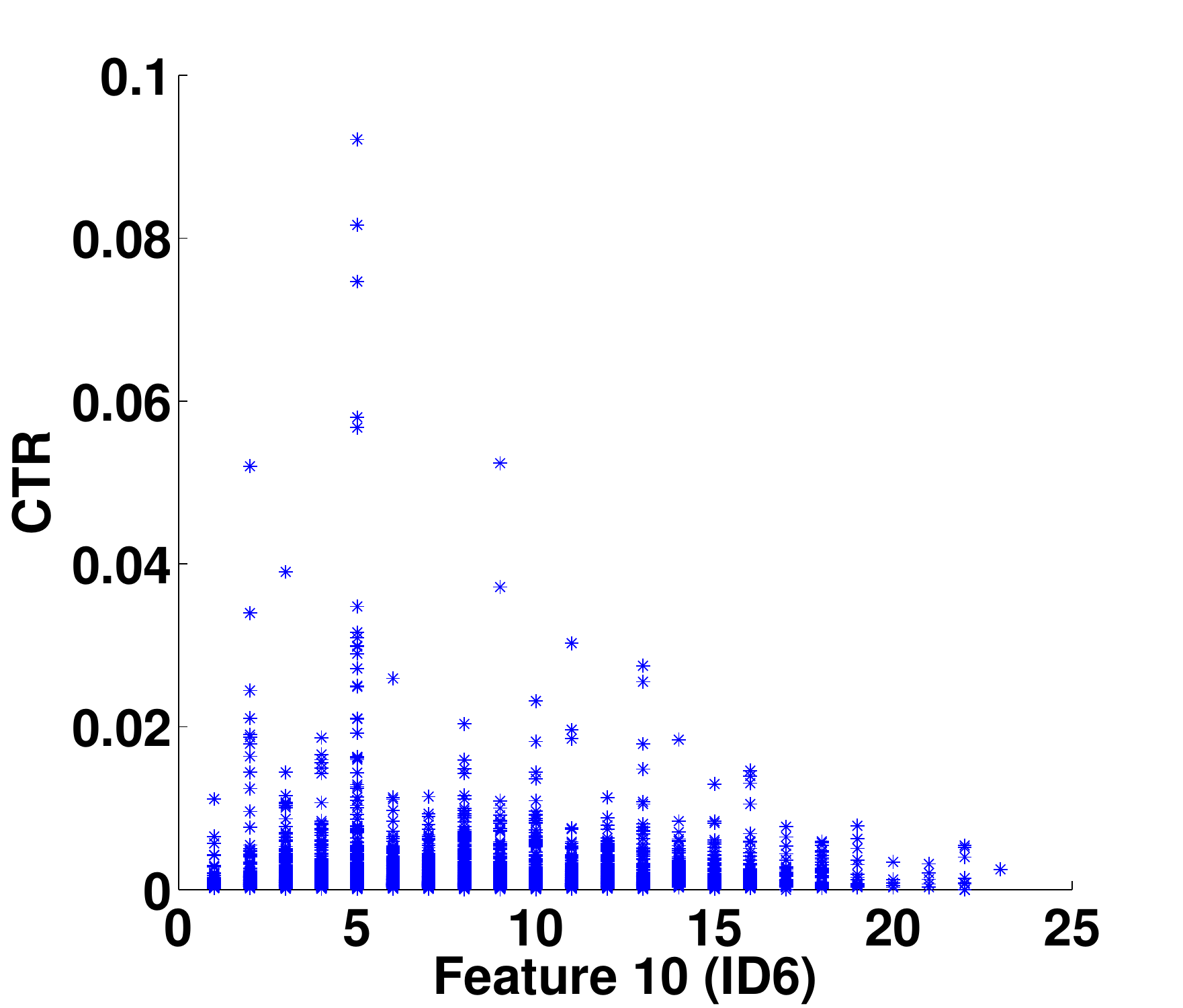}  \\
\end{tabular}
\caption{The scatter plot of $f_{10}$.}
\label{fig:s5-dis}
\end{center}
\end{figure}

$S_5$ is the best feature set for data set ID$2$ and the fourth for ID6. It only includes $f_{10}$ which is the number of connected coherent components. The scatter plot of $f_{10}$ in both data sets are shown in figure \ref{fig:s5-dis}. The scatter plot shows that creatives with more than $15$ connected coherent components in data set ID$6$ and more than $20$  in data set ID$2$ are unlikely to achieve a CTR higher than $0.01$. In other words, this suggests that cluttered creatives containing many objects tend to have lower CTR.

\begin{figure}
\begin{center}
\begin{tabular}{c c }
\includegraphics[width=1.5in, height=1.4in] {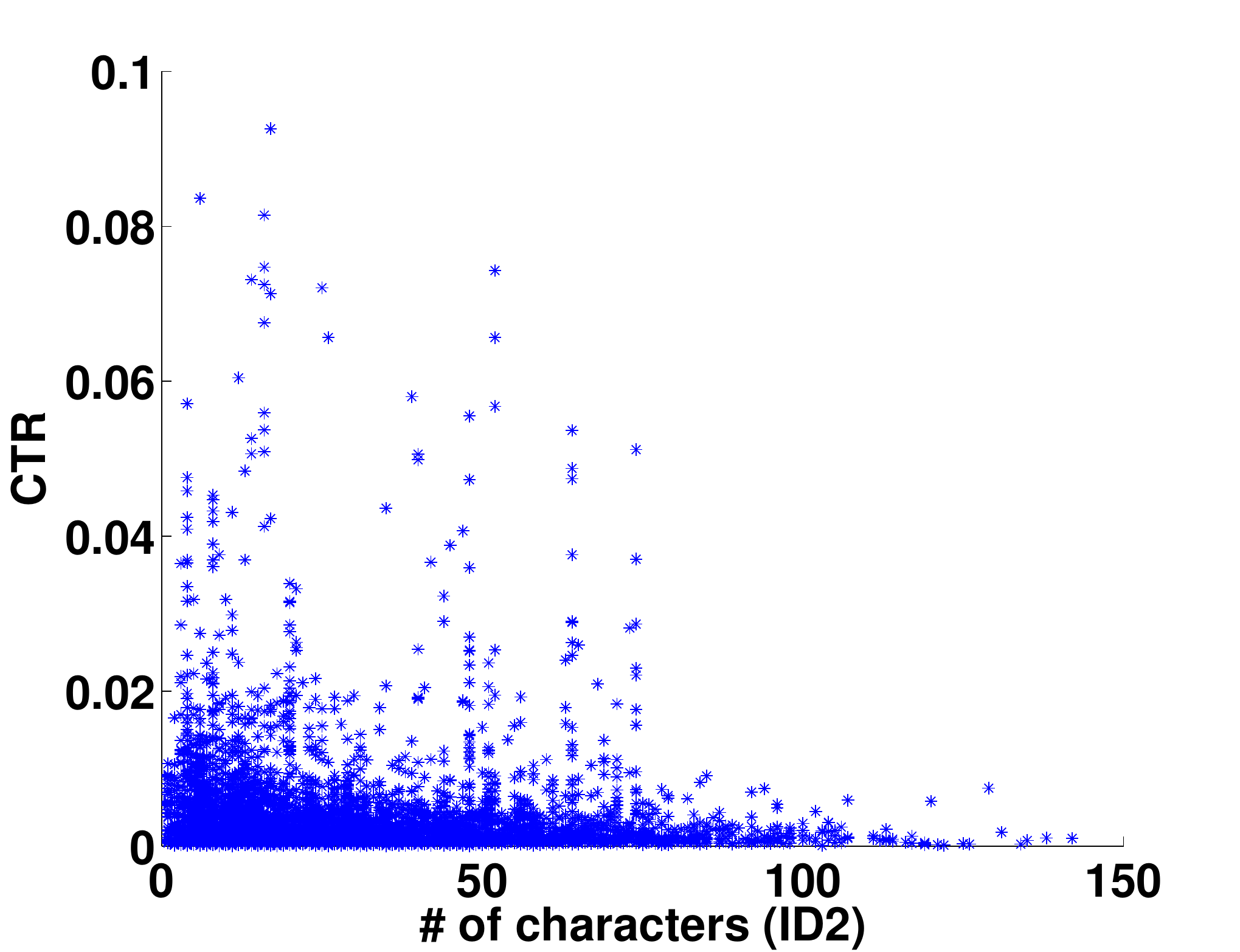} &
\includegraphics[width=1.5in, height=1.4in] {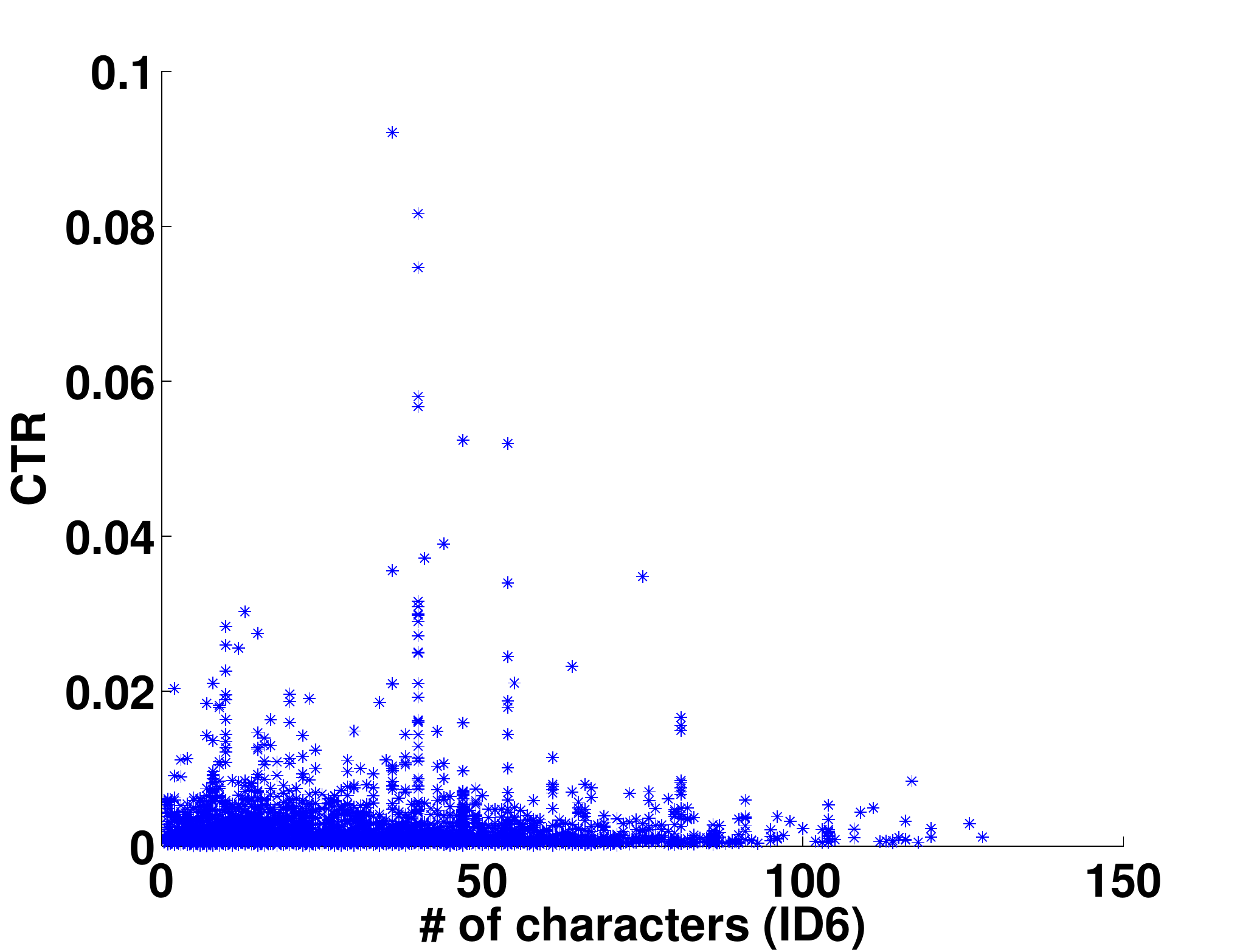}  \\
ID$2$ &    ID$6$ \\
\end{tabular}
\caption{The scatter plot of number of characters.}
\label{fig:ocr}
\end{center}
\end{figure}

The number of characters, $S_{19}$ in data set ID$2$ and $S_{20}$ in data set ID$6$, is interestingly the third important feature set in both data sets. Figure \ref{fig:ocr} shows the scatter plot of the number of characters in both data sets. It can be seen that the creatives with higher number of characters are unlikely to achieve high CTR values in both data sets, once again suggesting that textual clutter is undesirable.

The next selected categories is $S_{17}$ which is the $4$-th selected category in ID$2$ and the $5$-th in ID$6$. $S_{17}$ represents the number of connected components in saliency binary map, distance between salient components, distance of saliency areas from the center of image and rule of third closest point. This indicates the importance of saliency features as well as considering professional photography rules such as the rule of third in designing ads. Intuitively, a small number of salient components, closer to the center of the creative, and consistent with the rule of third are desirable features in a creative. Finally, $S_{13}$, which contains features describing the number of hues and the contrast of hues in the largest segment of the image, is the $5$-th important common category considering both data sets. Note that the scatter plot of the last $2$ selected categories have been omitted due to space limit. In summary, our top $5$ selected categories include the features from all proposed feature categories, global, local and advanced features, indicating the importance of each of them in predicting the creatives CTR.


\begin{table}
\caption{Complete list of visual features.}
\label{table:fs-results}
\begin{center}
\begin{small}
\begin{tabular}{@{} c @{} @{}c @{} @{\ }c c@{\ } c@{\ } c  @{\ }c @{ }c@{} }
\textbf{Feature} & \textbf{Short Description} & \textbf{LC ID$2$} &\textbf{MI ID$2$} &  \textbf{LC ID$6$} &\textbf{MI ID$6$} &\textbf{C-ID$2$} & \textbf{C-ID$6$} \\ \hline
$f_1$ & gray level contrast& -0.048 &    0.063 &    -0.102&    0.098 &  $S_{1}$ &  $S_{1}$  \\
$f_2$ & number of dominant bins in gray level histogram &  -0.107 &    $0.071$&    $\textbf{-0.137}$&    $\textbf{0.112}$& $S_{1}$&  $S_{1}$\\
$f_3$ & standard deviation of gray level images &  -0.063 &    0.061 &    -0.016 &    0.082 & $S_{2}$&  $S_{2}$\\
$f_4$ & number of dominant bins in RGB histogram &  -0.083 &    0.067 &    -0.074 &    0.078 & $S_{3}$&  $S_{3}$\\
$f_5$ & size of the dominant bin in RGB histogram &  0.078 &    \textbf{0.093} &    0.064 &    \textbf{0.104} & $S_{3}$&  $S_{3}$ \\
$f_6$ & number of dominant bins in HSV histogram &  -0.083 &    0.073 &    -0.078 &    0.092 & $S_{3}$&  $S_{3}$ \\
$f_7$ & size of the dominant bin in HSV histogram &  0.119 &    \textbf{0.096} &    0.085 &    0.094 &  $S_{3}$&  $S_{3}$ \\
$f_8$ & deviation from the best color harmony model&  -0.062 &    0.039 &    -0.040 &    0.052 &  $S_{4}$&  $S_{4}$ \\
$f_9$ & average deviation from the best two color harmony models&  -0.069 &    0.047 &    -0.046 &    0.063 & $S_{4}$&  $S_{4}$ \\
$f_{10}$ & number of connected coherent components &  0.094 &    0.062 &    -0.004 &    0.064 & $S_{5}$&  $S_{5}$ \\
$f_{11}$ & size of the largest connected coherent component&  0.102 &    \textbf{0.090} &    0.069 &    0.099 & $S_{3}$&  $S_{3}$ \\
$f_{12}$ & size of the second largest connected coherent component&  0.085 &    0.068 &    \textbf{0.315} &    \textbf{0.122} & $S_{6}$&  $S_{6}$ \\
$f_{13}$ & color size rank of the largest connected coherent component&  -0.002 &    0.010 &    -0.012 &    0.014 &  $S_{7}$&  $S_{7}$ \\
$f_{14}$ & color size rank of the second largest connected coherent component&  -0.046 &    0.020 &    -0.039 &    0.022 & $S_{8}$&  $S_{8}$ \\
$f_{15}$ & number of dominant hues&  -0.102 &    0.036 &    -0.119 &    0.045 & $S_{9}$&  $S_{9}$ \\
$f_{16}$ & contrast of  dominant hues&  -0.015 &    0.027 &    -0.015 &    0.046 & $S_{9}$&  $S_{9}$ \\
$f_{17}$ & standard deviation of hues&  0.010 &    \textbf{0.084} &    -0.039 &    \textbf{0.102} & $S_{10}$&  $S_{10}$ \\
$f_{18}$ & average lightness&  0.031 &    0.072 &    0.035 &    0.099 & $S_{11}$&  $S_{11}$ \\
$f_{19}$ & standard deviation of lightness&  -0.065 &    0.063 &    -0.086 &    0.084 & $S_{2}$&  $S_{2}$\\
\hline
$f_{20}$ & size of the Largest Segments (LS)&  -0.052 &    0.058 &    -0.080 &    0.081 & $S_{12}$&  $S_{12}$  \\
$f_{21}$ & segments size contrast&  -0.044 &    0.054 &    -0.122 &    0.092 & $S_{12}$&  $S_{12}$ \\
$f_{22}$ & number of image dominant hues in the LS&  -0.104 &    0.044 &    -0.068 &    0.029 & $S_{13}$&  $S_{13}$ \\
$f_{23}$ & number of dominant hues in the LS&  -0.102 &    0.032 &    -0.080 &    0.036 & $S_{13}$&  $S_{13}$ \\
$f_{24}$ & largest number of dominant hues in one segment&  -0.079 &    0.031 &    \textbf{-0.125} &    0.045 & $S_{14}$&  $S_{14}$ \\
$f_{25}$ & contrast of hues number among segments&  -0.006 &    0.027 &    -0.098 &    0.033 &  $S_{14}$&  $S_{14}$ \\
$f_{26}$ & contrast of hues in the LS&  -0.120 &    0.032 &    -0.085 &    0.036 & $S_{13}$&  $S_{13}$ \\
$f_{27}$ & standard deviation of hues contrast among segments&  -0.056 &    0.072 &    -0.073 &    0.098 & $S_{10}$&  $S_{10}$\\
$f_{28}$ & deviation from the best color harmony model for LS&  -0.047 &    0.030 &    -0.034 &    0.042 &  $S_{4}$&  $S_{4}$\\
$f_{29}$ & average deviation from the best two color harmony models for LS &  -0.055 &    0.037 &    -0.049 &    0.053 & $S_{4}$&  $S_{4}$\\
$f_{30}$ & average lightness in LS&  0.027 &    0.075 &    0.065 &    0.096 & $S_{11}$&  $S_{11}$ \\
$f_{31}$ & standard deviation of average lightness among the segments&  0.080 &    0.076 &    -0.049 &    0.091 & $S_{15}$&  $S_{15}$\\
$f_{32}$ & contrast of average lightness among the segments&  0.075 &    0.067 &    -0.027 &    0.081 & $S_{15}$&  $S_{15}$\\
\hline
$f_{33}$ & background size in Saliency Map (SM)&  \textbf{-0.165} &    0.082 &    \textbf{-0.175} &    \textbf{0.106} & $S_{16}$&  $S_{16}$\\
$f_{34}$ & number of connected components in SM&  -0.111 &    0.029 &    -0.062 &    0.088 & $S_{17}$&  $S_{17}$\\
$f_{35}$ & size of the largest connected components  in SM&  \textbf{0.132} &    0.075 &    0.036 &    0.073 & $S_{16}$&  $S_{18}$ \\
$f_{36}$ & average saliency weight of largest connected component&  \textbf{0.165} &    0.075 &    0.071 &    0.080 & $S_{16}$&  $S_{18}$\\
$f_{37}$ & number of connected components in image background SM&  -0.012 &    0.015 &    -0.035 &    0.038 & $S_{18}$&  $S_{19}$ \\
$f_{38}$ & size of the largest connected component of background in SM&  \textbf{-0.163} &    \textbf{0.091} &    \textbf{-0.145} &    0.052 & $S_{16}$&  $S_{16}$ \\
$f_{39}$ & distance between connected components in SM&  0.117 &    0.056 &    -0.027 &    0.072 & $S_{17}$&  $S_{17}$ \\
$f_{40}$ & distance from rule of third points in SM&  \textbf{0.126} &    0.063 &    0.010 &    0.086 & $S_{17}$&  $S_{17}$ \\
$f_{41}$ & distance from center of image in SM&  0.091 &    0.065 &    0.011 &    0.087 &  $S_{17}$&  $S_{17}$ \\
$f_{42}$ & number of characters (OCR)&  -0.082 &    0.057 &    -0.017 &    0.071 & $S_{19}$&  $S_{20}$ \\
$f_{43}$ & number of faces in the image&  -0.035 &    0.008 &    -0.024 &    0.011 & $S_{20}$&  $S_{21}$ \\
\end{tabular}
\end{small}
\end{center}
\end{table}

\section{Conclusion}
\label{sec:conclusion}
In this paper we investigated the relationship between the user response rate and the visual appearance of creatives in online display advertising. To the best of our knowledge, this is the first work in this area. We designed $43$ visual features for our experiments. We extracted the features from large scale data produced by the world's largest ad exchange system. We tested the utility of visual features in CTR prediction, ranking and classification. The experimental results demonstrate that our proposed framework is able to outperform baseline consistently, indicating the efficacy of visual features in predicting CTR. We also performed feature selection to select the top visual feature categories that have strongest importance for increasing CTR. The findings from this work will be useful for ads selection and developing visually appealing creatives with higher user response propensity in online display advertising.

\bibliographystyle{natbib}
\bibliography{DisplayXiv}

\end{document}